\documentclass[fleqn,usenatbib]{mnras}

\usepackage{newtxtext,newtxmath}

\usepackage[T1]{fontenc}
\usepackage{orcidlink}
\usepackage{gensymb}

\DeclareRobustCommand{\VAN}[3]{#2}
\let\VANthebibliography\thebibliography
\def\thebibliography{\DeclareRobustCommand{\VAN}[3]{##3}\VANthebibliography}

\usepackage{graphicx}	
\usepackage{amsmath}	
\usepackage{multirow}
\usepackage{tikz}
\usepackage{float}

\title[\textsc{Diagnosing Wind-Formed Emission Lines}]{A Diagnostic Kit for Optical Emission Lines Shaped by Accretion Disc Winds}

\author[A. G. W. Wallis et al.]{
Austen G. W. Wallis$^{\orcidlink{0000-0003-0770-9015}}$,$^{1}$\thanks{E-mail: A.Wallis@soton.ac.uk}
Christian Knigge$^{\orcidlink{0000-0002-1116-2553}}$,$^{1}$
James~H.~Matthews$^{\orcidlink{0000-0002-3493-7737}}$,$^{2}$
Knox S. Long$^{\orcidlink{0000-0002-4134-864X}}$,$^{3,4}$
Stuart A. Sim$^{\orcidlink{0000-0002-9774-1192}}$$^{5}$
\\
$^{1}$School of Physics and Astronomy, University of Southampton, Southampton, SO17 1BJ, UK\\
$^{2}$Department of Physics, Astrophysics, University of Oxford, Denys Wilkinson Building, Keble Road, Oxford, OX1 3RH, UK\\
$^{3}$Space Telescope Science Institute, 3700 San Martin Drive, Baltimore, MD, 21218, USA\\
$^{4}$Eureka Scientific Inc., 2542 Delmar Avenue, Suite 100, Oakland, CA, 94602-3017, USA\\
$^{5}$School of Mathematics and Physics, Queen's University Belfast, University Road, Belfast, BT7 1NN, UK\\
}

\date{\today}

\pubyear{\the\year{}}

\begin{document}
\label{firstpage}
\pagerange{\pageref{firstpage}--\pageref{lastpage}}
\maketitle

\begin{abstract}
Blueshifted absorption is the classic spectroscopic signature of an accretion disc wind in X-ray binaries and cataclysmic variables (CVs). However, outflows can also create pure emission lines, especially at optical wavelengths. Therefore, developing other outflow diagnostics for these types of lines is worthwhile. With this in mind, we construct a systematic grid of 3645 synthetic wind-formed $\mathrm{H\,\alpha}$ line profiles for CVs with the radiative transfer code \textsc{sirocco}. Our grid yields a variety of line shapes: symmetric, asymmetric, single‐ to quadruple‐peaked, and even P-Cygni profiles. About 20 per cent of these lines -- our `Gold' sample -- have strengths and widths consistent with observations. We use this grid to test a recently proposed method for identifying wind-formed emission lines based on deviations in the wing profile shape: the 'excess equivalent width diagnostic diagram'. We find that our Gold sample can preferentially populate the suggested `wind regions' of this diagram. However, the method is highly sensitive to the adopted definition of the line profile `wing'. Hence, we propose a refined definition based on the full-width at half maximum to improve the interpretability of the diagnostic diagram. Furthermore, we define an approximate scaling relation for the strengths of wind-formed CV emission lines in terms of the outflow parameters. This relation provides a fast way to assess whether -- and what kind of -- outflow can produce an observed emission line. All our wind-based models are open-source and we provide an easy-to-use web-based tool to browse our full set of $\mathrm{H\,\alpha}$ spectral profiles.
\end{abstract}



\section{Introduction}
\subsection{Outflow Signatures}
\label{sec:OutflowsIntro}
Blueshifted absorption, a component seen in many line profiles, including the P-Cygni profile, has long served as the ultimate smoking gun for line formation in outflows. In the context of accretion-powered astrophysical systems, outflows are usually thought to be driven from an accretion disc's surface. Examples of disc-accreting systems that present these spectroscopic signatures include young stellar objects (YSOs; \citep{beristain_helium_2001,edwards_probing_2006}), low-mass X-ray binaries (LMXBs; \citep{miller_accretion_2008,trueba_comprehensive_2019,segura_persistent_2022}), cataclysmic variables (CVs; \citep{heap_iue_1978,cordova_high-velocity_1982,holm_ultraviolet_1982,drew_inclination_1987,la_dous_catalogue_1990,hartley_testing_2002,prinja_fuse_2003}) and broad absorption line quasars (BALQSOs; \citep{weymann_comparisons_1991,hall_unusual_2002, filiz_ak_dependence_2014, mcgraw_quasar_2018}).

However, although all P-Cygni profiles indicate outflows, not all outflows produce P-Cygni profiles for all possible observers. More specifically, this particular profile shape is only inevitable for all viewing angles if three conditions are satisfied: (i) the outflow is spherically symmetric; (ii) the continuum source illuminating the outflow is spherically symmetric; (iii) line formation is solely due to resonance scattering. None of these conditions usually hold in the context of accretion disc winds. The outflow is likely to be approximately biconical; the primary continuum source is usually the accretion disc, and collisional excitation and/or recombination contribute significantly to line formation \citep{matthews_impact_2015,matthews_testing_2016,matthews_stratified_2020, matthews_disc_2016}. As a result, disc winds can easily produce pure emission line profiles, especially when viewed from particular orientations \citep{vitello_ultraviolet_1993,froning_observations_2004,groot_spectrophotometric_2004, matthews_impact_2015}.

With such a variety of possible physical outflow conditions, outflows can leave detectable imprints on a line profile's shape, such as slight asymmetries induced by the different optical depths presented along different lines of sight. Observations of CVs like V455 And \citep{tampo_disc_2024} and LMXBs like MAXI J1820+070 \citep{munoz-darias_hard-state_2019, mata_sanchez_hard-state_2022} vividly illustrate how short-timescale variations in disc winds properties can significantly alter these optical emission line shapes. However, identifying whether a particular profile shape signals line formation in an outflow is tricky. Historically, researchers have sometimes used the moments of a spectral line, such as skew and kurtosis, as measures of asymmetry to aid the diagnosis of a line-emitting gas's geometry and kinematics \citep{boroson_emission-line_1992, leighly_hst_2004}. However, spectral lines are not genuine probability distributions, making these measures typically quite challenging to implement and interpret. Furthermore, even a simple velocity shift of an emission line may result from disc-related features or an obscuration/suppression of flux in one wing, instead of originating from the true bulk motion of a gas \citep{prinja_high-resolution_1995, richards_broad_2002}. As a result, there is no foundational method for identifying and diagnosing wind-formed emission line profiles. 

\subsection{Line Wing Equivalent Width Excesses and Diagnostic Diagrams}
\label{sec:DiagPlots}

\begin{figure*}
    \centering
    \includegraphics[width=\textwidth]{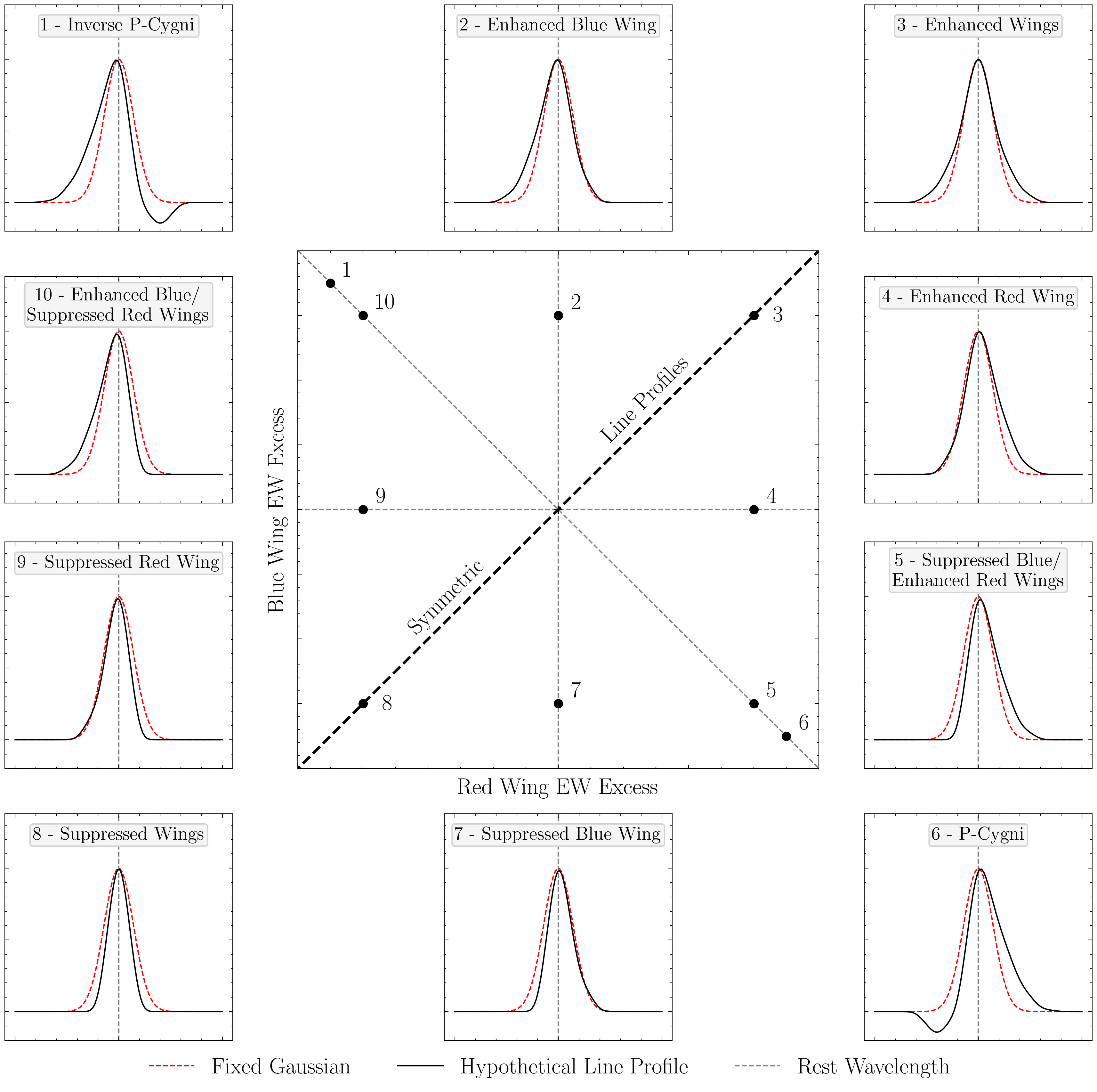}
    \caption{A qualitative, idealised sketch of various possible single-peaked line profiles numbered with their respective `expected' data point positions on an EW excess diagnostic diagram. The hand-drawn line profiles (solid black) are a small subset of many shape variations physically possible. However, when compared against the fixed Gaussian (dashed red) across all plots, these profiles highlight the excess EW value impact from shape deviations. Excess data points on the $\mathrm{y=x}$ line are line profiles with expected (but not guaranteed) symmetric deviations from the Gaussian. Other asymmetric deviations, that is, different deviations between the blue and red sides of the Gaussian mean or rest wavelength (dashed grey), populate regions away from this symmetric line. For example, we would expect the iconic P-Cygni profile to be characterised in a similar region to the line profile labelled `Suppressed Blue/Enhanced Red Wings'.}
    \label{fig:diag_example}
\end{figure*}

In an effort to address this lack of a go-to method, \citet{mata_sanchez_1989_2018} first introduced the `excess diagnostic diagram' -- a plot of blue vs red equivalent‐width (EW) excesses -- as a means to track wind‐related features during the 1989 and 2015 outbursts of the low‐mass X‐ray binary V404 Cygni. As explained in detail in this section, the method calculates the blue and red EW excesses associated with a given line profile relative to a Gaussian that best fits the spectral line’s core. Any broad, non‐Gaussian or asymmetric wings that remain after subtracting this Gaussian are then attributed to outflow signatures. The blue and red excesses quantify how much -- and on which side -- a spectral line deviates from a symmetrical Gaussian profile. We adopt a Gaussian simply because the function provides a well‐defined, symmetric reference shape and follows \citet{mata_sanchez_1989_2018}'s original methodology. Later, \citet{mata_sanchez_ask_2023} developed this method further as a tool for systematically detecting and classifying outflows.

To aid this conceptual description, Fig. \ref{fig:diag_example} shows a qualitative, idealised example of a `red-blue line wing EW excess diagnostic diagram'. Additionally, to help with the brevity of this naming, we simplify the terminology to variants of `EW excesses' throughout the rest of this paper. The central diagnostic diagram displays the calculated excess EWs for all fundamentally distinct cases, each illustrated by a neighbouring line profile. In each case, the calculated excesses correspond to differences, expressed in EWs, between a particular emission line profile (solid black) and a Gaussian (dashed red) aligned with the line's centre (positioned at the rest wavelength for this diagram). In calculating these excesses, the line is split into red and blue sides, separated by the line's rest wavelength. As a result, each line yields a red wing excess EW and a blue wing excess EW. These two parameters then define the x-y plane of the diagnostic diagram. The surrounding examples in Fig. \ref{fig:diag_example} intend to highlight how the location of a line profile in the diagnostic diagram connects to the actual appearance of the line profile. The figure also provides brief physical descriptions of these example profiles. Note that in Fig. \ref{fig:diag_example}, the Gaussian is identical in all examples. However, this choice only aims to help facilitate interpretation and build intuition about the diagnostic diagram until the methodological details in Section \ref{sec:method}. In practice, each line profile must have a Gaussian fit.

At first glance, EW excess diagrams can look confusing. However, the idea is that one can simply identify/classify complex observed line profiles by fitting a Gaussian. The data points' position in the diagram aids in understanding the underlying kinematics of the flow that produced the line profile. For example, consider the `suppressed blue/enhanced red wing' hypothetical line profile in the bottom right quadrant. This shape is essentially our `smoking gun' outflow signature, the P-Cygni profile. The suppressed emission or possible absorption component results in a negative blue wing excess; conversely, the enhanced red emission results in a positive red wing excess. The more extreme a deviation is from the Gaussian, the further a data point lies from the origin. Hence, these plots are typically only constructed for strong optical lines, such as $\mathrm{H\,\alpha}$, $\mathrm{H\,\beta}$ and $\mathrm{He\,\textsc{I}}$, where even the broad wings of the profiles can be confidently tracked. 

\subsection{Scope and Objectives}
\label{sec:Overview}

EW excess diagrams have already been applied in multiple studies to identify and study outflows in LMXBs and CVs \citep{mata_sanchez_1989_2018,munoz-darias_hard-state_2019,panizo-espinar_optical_2021,cuneo_unveiling_2023}. More recently, \citet{mata_sanchez_ask_2023} used these diagrams to validate a neural network-based machine learning tool designed to detect and classify outflow signatures in spectral line profiles. However, the diagnostic method itself has not yet been fully validated. In particular, with the possible exception of the lower right ('P-Cygni') quadrant, there is uncertainty where wind-formed lines are likely to lie predominantly within diagnostic diagrams. Moreover, specific methodological choices, notably the definition adopted for what constitutes a line's `wing', still require thorough evaluation.

This paper aims to address these unknowns and provide a simple procedure for estimating the strength of wind-formed $\mathrm{H\,\alpha}$ lines in CVs. Hence, we create an extensive and systematic sample of simulated disc wind $\mathrm{H\,\alpha}$ line profiles for CVs using \textsc{sirocco}, an advanced Monte-Carlo radiative transfer code for outflowing media \citep{matthews_sirocco_2024}. These simulations allow us to determine the effectiveness of the EW excess diagnostic diagram for line characteristics and provide an approximate calibrated scaling law relation that connects equivalent width to physical wind parameters.

We outline the paper as follows. Firstly, in Section \ref{sec:Sirocco}, we will briefly discuss the \textsc{sirocco} modelling code and explain our particular wind parameter choices in creating a systematic grid of physically plausible synthetic line profiles. Next, in Section \ref{sec:method}, we describe how excess EWs are calculated in practice, including the methodological choices that must be made. We also point out some minor modifications we have made to make the original method work more robustly for the wide range of line profile shapes produced by our simulations. In Section \ref{sec:Benchmark}, we introduce two observational data sets we use as a point of comparison for our simulations. Then, in Section \ref{sec:sirocco_lines}, provide an overview of the line profile shapes seen within our \textsc{sirocco} grid. We describe the filters applied to remove spectra for which the excess EWs cannot be meaningfully estimated and discuss further limits on our \textsc{sirocco} lines to ensure we only compare `realistic' profiles to our observational benchmark. In Section \ref{sec:results}, we present our excess diagnostic diagrams following the original methodology for outflowing systems at five different inclinations and highlight issues uncovered during the process. Subsequently, we discuss the implications of our findings in Section \ref{sec:discussion} and propose practical improvements utilising the FWHM that benefit the expected behaviour of the diagnostic. We also use our model grid to construct and calibrate a set of approximate scaling law relations to connect $\mathrm{H\,\alpha}$'s EWs and physical wind parameters. This relation provides a fast and effective starting point for interpreting and/or modelling observations. We summarise and conclude in Section \ref{sec:conclusion}.

\section{\textsc{sirocco} Outflow Simulations}
\label{sec:Sirocco}

\textsc{sirocco} is a Monte-Carlo ionization and radiative transfer code for modelling astrophysical outflows \citep{matthews_sirocco_2024}. The code is designed to simulate the spectra of astrophysical systems, given a set of user-defined physical parameters. A user can select different pre-defined models of azimuthally-symmetric outflows for a variety of systems, including AGN \citep{matthews_disc_2023}, CVs \citep{matthews_impact_2015} and tidal disruption events \citep{parkinson_multi-dimensional_2024}. The code assumes the astrophysical system is near radiative equilibrium and in a steady state. This assumption allows \textsc{sirocco} to iteratively compute the radiation field and ionization state of the outflow. In this `ionization cycle' stage, photon packets travel through and interact with the plasma, thus modifying the outflow's heating, cooling and ionization rates. Once the temperature and ionization structure have converged, \textsc{sirocco} freezes the plasma's state. Then, the `spectral cycle' stage begins, during which the code calculates the observed spectra. Here, the code emits photon packets from user-defined radiation sources and `observes' the light received along particular lines of sight after passing through the outflow.

For this study, we simulated 729 CV systems using wind parameters that cover a wide range of physically plausible configurations. In producing this grid of models, we first defined a set of default (or reference) physical parameters. Table \ref{tab:combination table} gives this complete list of parameters defining our models. We then construct our grid by defining `high' and `low' values on either side of the default values for 6 key parameters; these are highlighted in bold in Table \ref{tab:combination table}. This procedure results in a quasi-regular 6-dimensional grid containing $3^6 = 729$ models. The 6 parameters we vary for our grid space are:

\begin{enumerate}
    \item \textbf{The disc mass accretion rate} of a standard Shakura-Sunyaev disc. This parameter controls the disc temperature distribution \cite[see, for example,][]{matthews_sirocco_2024}. We adopt three values of $\mathrm{3\times10^{-9}}$, $\mathrm{1\times10^{-8}}$ and $\mathrm{3\times10^{-8}\,M_\odot\, yr^{-1}}$. These correspond to a typical high-state (wind-driving) CV with a low, medium and high accretion rate.
     \footnote{Dwarf novae in quiescence can have significantly lower accretion rates, but no outflow signatures have been observed in this state \cite[see, however,][]{scepi_turbulent_2018}}
    \item \textbf{The wind mass-loss rate} from the disc as a function of the accretion rate. This parameter controls the overall rate at which matter is ejected across the accretion disc. We adopt three values of $0.03\times$, $0.1\times$ and $0.3\times$ the disc mass accretion rate. As the multiple increases, the system ejects an increasingly higher fraction of the disc's accreting material.
    \item \textbf{The wind's degree of geometric collimation [d]} as described by the Knigge--Wood--Drew biconical wind model in section 3.2.2 of \citet{knigge_application_1995}. \textsc{sirocco} implements the d parameter in units of the central object's radius, a white dwarf (WD) for a CV system, to generate the wind cone's inner and outer opening angles following $\theta_{X} = \tan^{-1}\left(\frac{r_{disc_{X}}}{d}\right)$, where $\theta_{X}$ is the angle from the disc's rotational axis, $r_{disc_{X}}$ is the disc's radius in $R_{WD}$ units and $X = min \ \mathrm{or} \ max$, the minimum or maximum bounds respectively. We adopt three values of 0.55, 5.5 and $55\,R_{WD}$ for the d component of the geometry parameter. The lower value defines an equatorial wind shape constrained to $61-89^{\circ}$. The middle value describes a much less collimated wind with opening angles spanning $10-80^{\circ}$. The high value produces a highly collimated outflow with opening angles in the $1-29^{\circ}$ range.
    \item \textbf{The [$\mathrm{\alpha}$] exponent controlling the specific wind mass-loss rate [$\dot{m}$]}, given the local flux/temperature of the disc. This parameter is defined via $\dot{m}\propto F(R)^{\alpha} \propto T(R)^{4\alpha}$, where $T$ and $F$ are the local effective temperature and associated flux, respectively. We adopt three values of 0, 0.25 and 1 for $\alpha$. These values correspond to a roughly constant specific mass-loss rate, a specific mass-loss rate proportional to temperature and a specific mass-loss rate proportional to flux across the disc.
    \item \textbf{The acceleration length scale [$R_s$]} of the outflow. The wind velocity law we use, given by equation 17 in \citet{knigge_application_1995}, features two control parameters that determine how the wind velocity increases as a function of distance along a given streamline. The acceleration length scale is one of these. We adopt three values of $1$, $10$ and $100\,R_{WD}$ for $R_s$. Longer acceleration length scales correspond to more gradual acceleration. In other words, slower velocities at any given distance along a streamline. 
\begin{table}
    \centering
    \begin{tabular}{l l}
        \hline
        \textbf{Parameter* [symbol] (unit)} & \textbf{Value} \\
        \hline
        System type & CV\\
        Central object mass ($\mathrm{M_\odot}$) & 0.8 \\
        Central object radius (cm) & $7.25182 \times 10^8$ \\
        Central object radiation & No \\
        Binary secondary mass ($\mathrm{M_\odot}$) & 0.4 \\
        Binary period (h) & 3.2 \\
        Disc type & Flat \\
        Disc radiation & Yes \\
        Disc radiation type & Blackbody \\
        Disc temperature profile & Standard \\
        \textbf{Disc mass accretion rate $\mathrm{(M_\odot\,yr^{-1})}$} & $\mathbf{3\times10^{-9}}$, $\mathbf{1 \times 10^{-8}}$, $\mathbf{3 \times 10^{-8}}$\\
        Disc maximum radius (cm) & $2.17555 \times 10^{10}$ \\
        Boundary layer radiation & No \\
        Wind number of components & 1 \\
        Wind type & Knigge-Wood-Drew \\
        Wind coordinate system & Cylindrical \\
        Wind dimension in x-direction & 30 \\
        Wind dimension in z-direction & 30 \\
        Photons per cycle & $10^7$ \\
        Ionization cycles & 25 \\
        Spectrum cycles & 220 \\
        & [440 for wider $\mathrm{\lambda}$ bounds]\\
        Wind ionization & Matrix\_pow \\
        Line transfer method & Macro\_atoms\_thermal\_trapping \\
        Atom transition mode & Matrix \\
        Atomic data file & data/h20\_hetop\_standard80.dat \\
        Surface reflection or absorption & Reflect \\
        Wind heating processes & Adiabatic \\
        \textbf{Wind mass-loss rate (in $\dot{\mathrm{M}}_{\mathrm{disc}}$)} & \textbf{0.03, 0.1, 0.3}\\
        \textbf{KWD Collimation [d] (in $\mathrm{R_{WD}}$)} & \textbf{0.55, 5.5, 55} \\
        \textbf{KWD Mass loss rate exponent [$\mathrm{\mathbf{\alpha}}$]} & \textbf{0, 0.25, 1.0}\\
        KWD Velocity factor [$\mathrm{v_\infty}]$ (in $\mathrm{v_{\text{escape}}}$) & 2.0 \\
        \textbf{KWD Acceleration length scale [$R_s$] (cm)} & $\mathbf{7.25182 \times 10^{8},}$\\
        & $\mathbf{7.25182 \times 10^{9},}$\\
        & $\mathbf{7.25182 \times 10^{10}}$ \\
        \textbf{KWD Acceleration exponent} [$\beta$] & \textbf{0.5, 1.5, 4.5} \\
        KWD Initial velocity [$\mathrm{v_0}$] (in sound speed) & 1.0 \\
        KWD Inner boundary [$\mathrm{r_{\text{min}}}$] (in $\mathrm{R_{WD}}$) & 1.0 \\
        KWD Outer boundary [$\mathrm{r_{\text{max}}}$] (in $\mathrm{R_{WD}}$) & 30.0 \\
        Wind maximum radius from CV (cm) & $10^{12}$ \\
        Wind initial temperature (K) & 40000.0 \\
        Wind filling factor & 1.0 (smooth)\\
        Disc radiation type in final spectrum & Blackbody \\
        Spectrum wavelength range (Å) & 6385 -- 6735 \\
        & [6210 -- 6910 wider bounds]\\
        Number of observers & 5 \\
        Spectrum angles / Inclination ($\mathrm{^{\circ}}$) & 20, 45, 60, 72.5, 85.0 \\
        Spectrum orbit phase & 0.5 \\
        Spectrum extraction method [live-or-die]  & extract\\
        Spectrum type  &  $\mathrm{\lambda F(\lambda)}$\\
        Reverb type & None \\
        Photon sampling approach & CV\\
        \hline
        \multicolumn{2}{| l |}{*NB: Some parameter labels have been altered to provide short descriptive}\\
        \multicolumn{2}{| l |}{names. See the documentation for full parameter descriptions and exact labels.}
    \end{tabular}
     \caption{The default parameters used for the simulation grid. This list is a template .pf file typically seen when simulating with \textsc{sirocco}. The regular font parameters are fixed values across all simulations. The bold font parameters are the physical parameter values we vary to create a regular grid of simulations. Each varying parameter has three corresponding grid point values, leading to $\mathrm{3^6}$ possibilities and 729 total spectra within our grid. Note that for the wind mass-loss rate, the parameter value is a multiple of the disc mass accretion rate for that particular simulation. We use a Knigge--Wood--Drew (KWD) wind model for the outflow \citep{knigge_application_1995}. For more detailed information on a particular parameter, see the \textsc{sirocco} \href{https://sirocco-rt.readthedocs.io/en/latest/input.html}{documentation}.}
     \label{tab:combination table}
\end{table}

    \item \textbf{The wind velocity law's acceleration exponent [$\mathrm{\beta}$]} is the other velocity law control parameter. We adopt three values of 0.5, 1.5 and 4.5 for $\mathrm{\beta}$. Larger $\mathrm{\beta}$ corresponds to slower acceleration. The effect of $\beta$ is particularly pronounced near the base of the wind. For large values of $\beta$, velocities remain well below the terminal velocity, $v_\infty$, out to streamline distances of $\simeq R_s$. This can effectively produce a dense `transition region' between the disc and the faster-moving parts of the outflow \cite[e.g.][]{knigge_disks_1997, knigge_eclipse_1997}.
\end{enumerate}

We run 25 ionization cycles for each simulation with 10 million photons per cycle. This number of iterations typically results in approximately 80 to 90 per cent of grid cells converging to equilibria within the outflow. Within the \textsc{sirocco} collaboration, we consider these percentages as the point when outflows have reached a steady state. Further ionization cycles beyond this point do not usually improve the convergence percentage. For spectral cycles, the total number of photons passing through the outflow is directly linked to the signal-to-noise ratio of a spectrum, where typically $S/N \propto \sqrt{N}$. Each model is computed with 2.2 billion photon packets when we adopt the narrow spectral window ($6385 - 6735$ Å), and with 4.4 billion packets when we use the broader window ($6210 - 6910$ Å); the latter is doubled to preserve a similar signal-to-noise (packet count) per spectral bin. This choice ensures that the statistical errors on our model EW excesses are comparable to those inferred from the observations reported by \citet{cuneo_unveiling_2023}. We chose five observer inclinations ($\mathrm{i}$) at 20°, 45°, 60°, 72.5° and 85° so that their $\mathrm{cos(i)}$ values are approximately uniformly distributed.

\section{EW Excess Diagnostic Diagrams: Implementation and Interpretation}
\label{sec:method}

The basic methodology for calculating EW excesses and constructing the associated diagnostic diagrams has already been described by \citet{mata_sanchez_1989_2018, cuneo_unveiling_2023}. However, these papers, as well as several other studies \cite[including, but not limited to][]{munoz-darias_hard-state_2019,munoz-darias_changing-look_2020,cuneo_discovery_2020, panizo-espinar_optical_2021} have typically implemented the methodology with slight, but potentially significant differences in fitting techniques and masking profiles. Since such differences \textit{could} introduce systematic uncertainties when interpreting the results, we will fully describe our understanding and implementation of the methodology here for clarity.

\subsection{The Original Methodology}
The original method requires 4 steps to construct an excess EW diagnostic diagram from a set of observed (or simulated) spectra. Fig. \ref{fig:masking_method} provides a visual aid for steps 1 to 3:
\begin{enumerate}
    \item A polynomial is fitted to each spectrum's continuum. Depending on the wavelength range of the spectrum, a mask may be applied over spectral lines. This mask helps to ensure that the fit only applies to regions of the spectrum that are continuum-dominated. The spectrum is then normalised, in other words, divided by the fitted polynomial, fixing the continuum level to a value of 1.
    
    \item A Gaussian is fitted to the continuum-normalised spectrum's emission line of interest. The Gaussian's amplitude, standard deviation and mean are free parameters. Since the idea is that non-Gaussian outflow signatures will primarily affect the line {\em wings}, the fit is constrained to a wavelength region centred on the line's core, masking out the wings. In the original method, this masking window is defined by the user and may change depending on the source or application. For example, \citet{cuneo_unveiling_2023} used $\mathrm{\pm\, 1000\,km\,s^{-1}}$.

\begin{figure}
    \centering
    \includegraphics[width=1.0\linewidth]{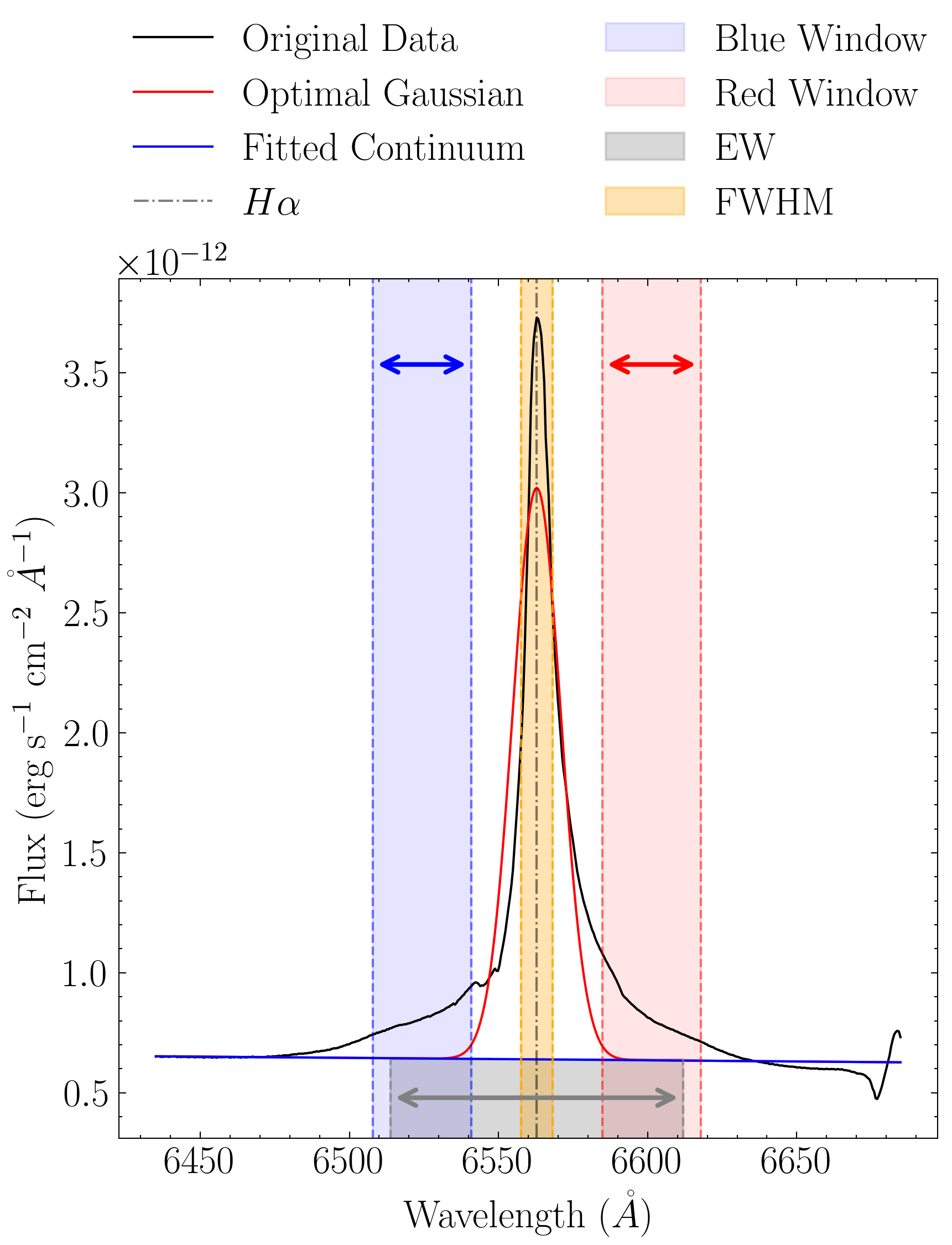}
    \caption{An example \textsc{sirocco} line profile fit following our modified excess diagnostic diagram methodology (see Section~\ref{sec:modifications}). The black line shows one particular simulation's $\mathrm{H\,\alpha}$ emission line within our grid space. The red line indicates the best-fitting Gaussian obtained by minimizing the least-squares residuals to the line profile, shown overlaid on the fitted continuum in blue. The grey dashed line indicates the rest wavelength of $\mathrm{H\,\alpha}$ at $\mathrm{6562.819\,\mathring{A}}$. The blue and red shaded regions show the user-chosen masking window in which we calculate the EW excess values. The window regions can vary in bounds depending on the user's decision. The shaded grey region shows the rectangular width of an area that extends from 0 to continuum equal to the spectral line's EW. The shaded orange region highlights the line's FWHM. Note the relatively poor fitting of the Gaussian for this non-Gaussian line with narrow core and broad wings, even though its residuals are optimally minimised. We discuss the limitations of this fitting procedure and impact on other line profile types more in Section~\ref{sec:modifications}, ~\ref{sec:H_alpha profiles} and ~\ref{sec:Filtering}.}
    \label{fig:masking_method}
\end{figure}
    \item The Gaussian fit is subtracted from the normalised spectrum to give the asymmetric excess emission. The excess EW is then calculated by integrating across the spectrum's blue and red wings. A masking window is again applied to this calculation to ensure that only high-velocity components are included. That is, the core is excluded, since the Gaussian may not fit the core region well, and/or shifts in the core may not be a sole consequence of outflow behaviour. Across studies, there has been a great deal of variation in how this masking window is defined. However, the lower bound should correspond to the value previously used for the Gaussian core fit. The upper bound should be set to a value where the spectral line seems to merge with the continuum emission. For example, \citet{cuneo_unveiling_2023}'s window has been defined as both $\mathrm{\pm\, 1000-2500\,km\,s^{-1}}$ and $\mathrm{\pm\, 400-1500\,km\,s^{-1}}$. In this paper's analysis, we adopt a $\mathrm{\pm\, 1000-2500\,km\,s^{-1}}$ masking window for observational comparison to  \citet{cuneo_unveiling_2023}. This step can be incorporated as a single calculation following
    \begin{equation}
    \label{equ:EW_Excess}
        EW_{\mathrm{Excess}} = \sum^{v_{\mathrm{high}}}_{v_{\mathrm{low}}, i} \frac{F_{v_i} - F_{G_i}}{F_{c_i}} \Delta v_i
    \end{equation}
    where $F_{v}$ is the spectrum,  $F_{G}$ is the Gaussian fit, $F_{c}$ is the continuum fit (assumed 1 if using a normalised spectrum), $\Delta v$ is the radial velocity width of the corresponding spectral wavelength bin, $v_{\mathrm{high}}$ defines the high radial velocity masking window bound and $v_{\mathrm{low}}$ the low bound. Two separate instances of this equation are applied for the blue and red sides of the spectral line's rest wavelength. Excess emission is defined as positive for diagnostic diagrams.

    \item Finally, the EW excess values can be plotted on a diagnostic diagram with red excess on the x-axis and blue excess on the y-axis. If available, flux errors are carried through traditional error propagation methods and/or the spectra can be bootstrapped to generate errors on the excesses.
    
\end{enumerate}

\subsection{Modifications to the Original Method}
\label{sec:modifications}

This paper aims to closely follow the original method stated above, but with some slight modifications to accommodate simulated spectra and the findings of this study. Firstly, due to the expensive computational cost associated with generating high signal-to-noise \textsc{sirocco} spectra, this study solely fits $\mathrm{H\,\alpha}$ within a narrow wavelength band. Therefore, we fit a linear underlying continuum instead of the higher-order polynomial that was adopted in \citet{cuneo_unveiling_2023} due to their wider wavelength range. Similarly, we only fit the extreme wavelengths of our band to avoid contamination by the spectral line. We set these fitting window bounds to $\mathrm{6450-6475\,\mathring{A}}$ and $\mathrm{6625-6650\,\mathring{A}}$ corresponding to radial velocities at approximately $-5150$ to $\mathrm{-4000\,km\,s^{-1}}$ and $2850$ to $\mathrm{4000\,km\,s^{-1}}$, respectively. Broader lines with a wider spectrum wavelength range have different fitting window bounds. For these, we use $\mathrm{6300-6325\,\mathring{A}}$ and $\mathrm{6775-6800\,\mathring{A}}$, corresponding to radial velocities at approximately $-12000$ to $\mathrm{-10850\,km\,s^{-1}}$ and $9500$ to $\mathrm{10850\,km\,s^{-1}}$, respectively. The slight differences between the blue and red sides are to avoid contamination from the $\mathrm{He\,\textsc{I}-\lambda6678}$ line.

Secondly, we fix the radial velocity offset component, also known as the mean of the Gaussian fit, to the rest wavelength. As our simulated system is stationary and exclusively outflow-driven, any offset in the line's peak results from the wind. Hence, the asymmetric profile should encapsulate this skew within our EW excesses. When applying this choice to \citet{cuneo_unveiling_2023}'s observed spectra, there is no quantifiable difference except for an EW excess shift of $\approx 0.05\,\mathring{A}$ in BZ Cam's double-peaked spectra. However, observationally, the argument remains: how does one know if this shift is caused by a wind or something different? This is a tricky question with no real definitive answer. Spectra presenting widely separated double peaks or strongly skewed P-Cygni-like single peaks are naturally the most affected. However, we find that for \textit{most} observational and synthetic spectra, Gaussians with fixed means yield diagnostic plots that are more reflective of the actual line profile characteristics. Additionally, fixing the mean makes sense for our \textsc{sirocco} spectra, given that any offset or skew is guaranteed from the wind.

Thirdly, we relax the condition that fits the Gaussian to only the line's core. When applying this choice to the observed spectra in \citet{cuneo_unveiling_2023}, the EW excesses only shift within the expected ranges quantified by the original data point's error bars, essentially creating a negligible difference to the overall diagnostic diagram's data point scatter. However, our simulated spectra present a much wider range of line profile shapes, some of which have very broad non-Gaussian cores. Restricting the optimisation to a pre-defined `core' region can lead to very poor fits across the entire line in these cases. We obtain more reliable and meaningful excess EW results for all spectra by simply fitting the overall line profile. This change does mean we can worsen the fitting of the Gaussian to the line's core for an improvement in the tails for some profiles, as shown in Figure~\ref{fig:masking_method}. However, an important point to remember is that the methodology masks out this core region during the excess calculation, minimising the poor fitting core's overall impact and importance on the final excess EW results.

Finally, several observational studies augment their excess diagnostic diagrams with \textit{significance contours} as described by \citet{munoz-darias_hard-state_2019,panizo-espinar_optical_2021}. These contours are obtained by repeatedly measuring EW excesses with the same width masking window over pure continuum sections of the spectra. The resulting continuum residuals form an approximately zero-mean Gaussian, where $\sigma$ contours are derived. However, we do not produce significance contours as, in principle, they do not generate any new information beyond our bootstrapped error bars. By bootstrapping, the homogenous continuum noise is effectively encapsulated by the error bar, meaning the value derived from each excess divided by its error is essentially equivalent to the data point's contour `significance'. In practice, there might be some slight differences due to heteroscedastic errors, fitting systematics and systematic differences between continuum and line flux errors. Nevertheless, there is no expectation that these are the dominant source of error. For our simulated spectra, we resample our data 150 times. Given the signal-to-noise of the spectra, this yields errors typically less than $\mathrm{0.1\,\mathring{A}}$, which is similar in scale to errors found in \citet{cuneo_unveiling_2023}.

\section{Observational Benchmarks}
\label{sec:Benchmark}
\begin{figure*}
    \centering
    \includegraphics[width=\textwidth]{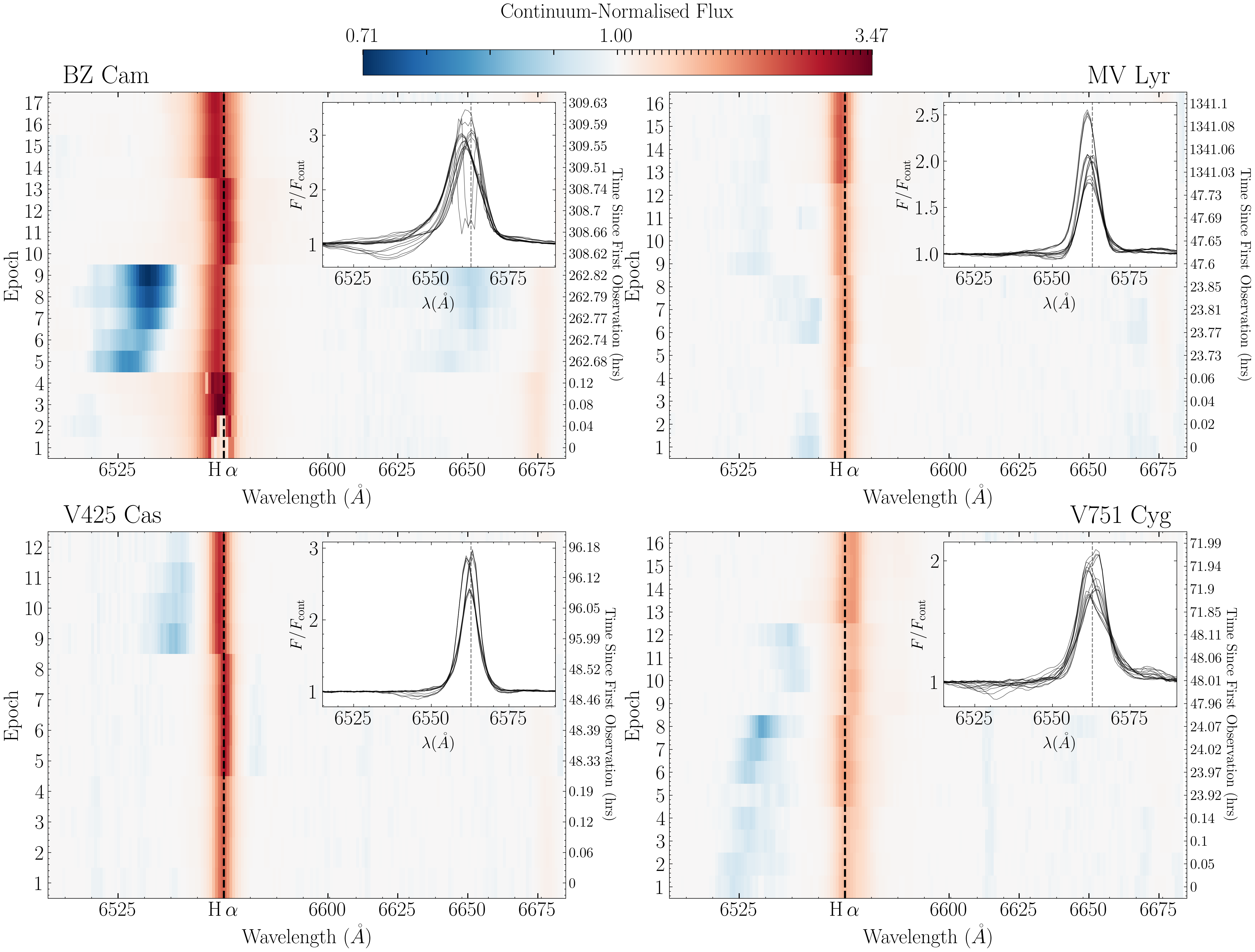}
    \caption{Trailed spectra of the $\mathrm{H\,\alpha}$ spectral line of four non-magnetic nova-like cataclysmic variable sources -- BZ Cam, V751 Cyg, MV Lyr and V425 Cas -- observed at different epochs. Redder colours indicate higher emission above the continuum, while bluer colours represent lower emission. The dashed black line highlights the rest wavelength of $\mathrm{H\,\alpha}$. Within each source's panel is an overlaid spaghetti plot of all the spectral $\mathrm{H\,\alpha}$ lines. These inset plots give an alternative view of the dataset's limited range of line profile shapes. The authors of \citet{cuneo_discovery_2020} note that the early epochs of BZ Cam suffer from line core saturation issues.}
    \label{fig:cuneo_line_examples}
\end{figure*}

The primary benchmark and point of comparison for our wind-formed line profiles is the observational data set used by \citet{cuneo_unveiling_2023}. This data set, shown in Fig. \ref{fig:cuneo_line_examples}, contains 61 observations of four non-magnetic nova-like cataclysmic variables -- BZ Cam, V751 Cyg, MV Lyr and V425 Cas -- all observed during normal high states. Even though the luminosity of these systems is dominated by their accretion discs, 58 of the 61 line spectra observed are strong, narrow and single-peaked. Approximately half of the spectra exhibit slight shifts of the emission line peak from rest, which the authors suggest is primarily caused by low-velocity absorption features, and most show asymmetric profiles from slight differences in high-velocity features between the blue and red wings. The lack of the classic double-peaked profile shape expected for disc-formed emission lines may at least partly be due to the relatively low inclination of these systems (estimated between $\mathrm{\approx10^{\circ}}$ to $\mathrm{50^{\circ}}$ \citep{skillman_superhumps_1995,ringwald_high-speed_1998,greiner_transient_1999,patterson_superhumps_2001,ritter_catalogue_2003, linnell_mv_2005, honeycutt_wind_2013}). 

The lead author of \citet{cuneo_unveiling_2023} kindly provided us with their reduced spectra for these systems. Given the slight modifications we have made to their method (as previously described in Section \ref{sec:modifications}), we have opted to calculate our own estimates of excess EWs for this data set, based on our revised methodology. However, our revised estimates typically agree with \citet{cuneo_unveiling_2023} within their respective uncertainties.

As showcased in the following section, our models produce a wide variety of line profile shapes, from essentially non-existent to much stronger than observed. This range is by design since we want to ensure our models span a large portion of the physically interesting parameter space. However, this diversity also means that some filtering is necessary to avoid producing diagnostic diagrams in which most simulated data points correspond to completely unrealistic models. We describe this filtering process in Section \ref{sec:Filtering}, but the final selection of our most realistic models, the `Gold' sample, enforces limits on each profile's EW and overall FWHM. In order to estimate these required limits, we also estimate the EW and FWHM for each of the profiles provided by \citet{cuneo_unveiling_2023}.  The EWs (in wavelength units) are calculated as 
\begin{equation}
    \label{equ:EW}
    EW = \sum^{\lambda_{\mathrm{high}}}_{\lambda_{\mathrm{low}}, i} \left ( \frac{F_{\lambda_i}}{F_{c_i}}-1 \right) \Delta \lambda_i.
\end{equation}
The FWHM is estimated by first determining the peak line flux value in the normalized spectrum, $F_{max}$. We then identify the bluest and reddest wavelengths where the flux first rises above the value $(\frac{F_{max}-1}{2})+1$. Here, $1 = F_{continuum}$. The difference between these wavelengths is recorded as our estimate of the FWHM.

We supplement this set of observed line strengths and widths with data from \citet{zhao_searching_2025}: a study based on low and medium-resolution spectra of accreting compact binaries by LAMOST. We extract 144 low-resolution and 113 medium-resolution data points for the EW and FWHM of CV $\mathrm{H\,\alpha}$ lines directly from fig. 2 of their paper. Two important points are worth noting here. 

First, \citet{zhao_searching_2025} estimates both EW and FWHM from Gaussian fits to the line profiles. Since this is a different method than the one we use to extract EW and FWHM from our simulated data and \citet{cuneo_unveiling_2023}'s spectra, some systematic differences may be expected between these estimates. Second, the \citet{zhao_searching_2025} sample includes all kinds of CVs, including both magnetic and non-magnetic systems. Among the non-magnetic systems, the sample also contains both dwarf novae (most of which would have been observed in quiescence, where no wind signatures are typically seen) and nova-like variables.

\section{\textsc{Sirocco}'s Grid of Wind-Formed $\mathrm{H\,\alpha}$ Lines}
\label{sec:sirocco_lines}
\subsection{Representative Spectral Examples}
\label{sec:H_alpha profiles}

\begin{figure*}
    \centering
    \includegraphics[width=\textwidth]{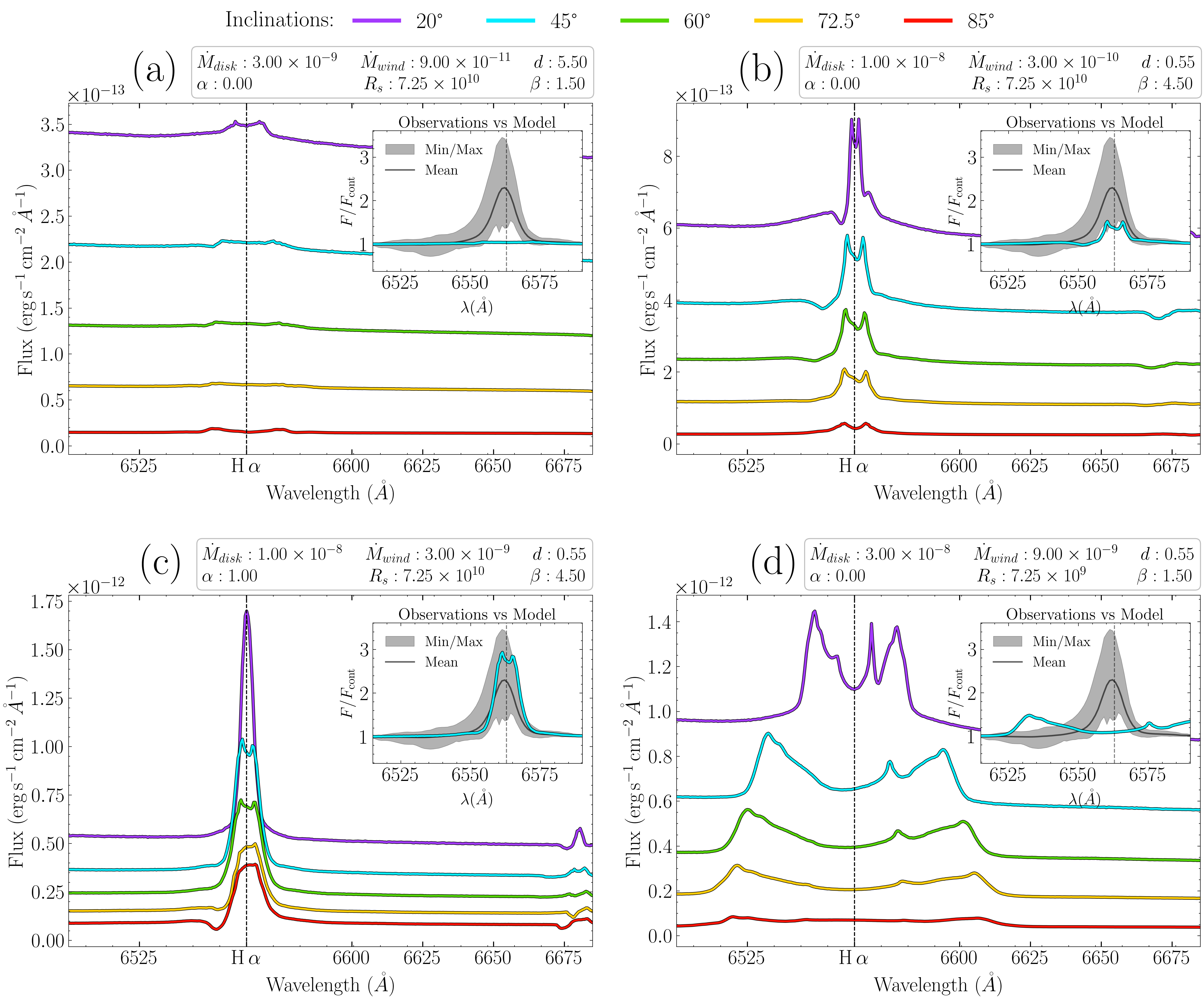}
    \caption{Four different \textsc{sirocco} simulations of $\mathrm{H\,\alpha}$ line profiles at $\mathrm{\approx6560\,\mathring{A}}$\ across inclinations from 20° to 85°. Plot (a) is typical of a weak $\mathrm{H\,\alpha}$ line profile. Plot (b) displays a P-Cygni $\mathrm{H\,\alpha}$ line profile with hints of an additional $\mathrm{He\,\textsc{I}-\lambda6678}$ line. Plot (c) shows a high flux $\mathrm{H\,\alpha}$ line profile with a noticeable but weaker $\mathrm{He\,\textsc{I}-\lambda6678}$ line. Plot (d) illustrates a more unusual multi-peak $\mathrm{H\,\alpha}$ line profile, which we ignore during our analysis. Each panel includes an inset plot comparing that model’s 45° inclination spectrum (cyan line) against our observational comparison (previously seen in Fig. \ref{fig:cuneo_line_examples}). The shaded grey region represents the full span of the observed line profiles (minimum to maximum flux) across all four CV systems, and the black line traces the mean of the 61 observed spectra. A \href{https://austenwallis.pyscriptapps.com/h-alpha-grid-inspector/latest/}{web-based interactive tool} is available to browse all our $\mathrm{H\,\alpha}$ spectral profiles and explore the impact of individual parameters.}
    \label{fig:sirocco_line_profiles}
\end{figure*}

Fig. \ref{fig:sirocco_line_profiles} illustrates four examples of the wide variety of wind-formed line profiles across our model grid and how different parameters influence the overall shape. Our full grid of line profiles, available to browse interactively (see Data Availability for details), are physically self-consistent wind-formed profiles produced by our specific outflow models. These spectra advance upon the simulated profiles created by \citet{mata_sanchez_ask_2023} for testing the EW excess method that previously relied on emulated line features. 

\textbf{\textit{Panel (a)}} shows an example of a set of weak line profiles with very little $\mathrm{H\,\alpha}$ emission. These types of profiles vary from continuum-only spectra to emission lines with minuscule peaks. Low accretion rates, wind mass-loss rates and acceleration exponents often define these types of profiles. When passing these profiles through a diagnostic diagram, the Gaussian fitting frequently struggles to fit the weak lines accurately. Additionally, as discussed in Section \ref{sec:filter sample} below, these line profiles do not resemble observations, so we remove most of these profile types from our analysis.

\textbf{\textit{Panel (b)}} shows a model producing more pronounced, but still weak, narrow double-peaked emission lines. This line also shows evidence of a P-Cygni profile at particular inclinations. These profiles can vary significantly in the strength and width of the double peak. We usually find these profiles at low to moderate wind mass-loss rates, but higher acceleration exponents and longer acceleration lengths. The wind collimation parameter primarily affects the peak-to-peak separation. Naturally, double-peaked lines are not well described by a Gaussian, even in the core. These lines are not too problematic if the peaks are narrow, as we only measure excesses within the masking window. However, if the peaks broaden and move into the masking window bounds, the two peaks will lead to a substantial and equal blue and red excess EW values. Such spectra would populate along the $\mathrm{y=x}$ diagonal of the diagnostic diagrams, as discussed further in section \ref{sec:results}. 

\textbf{\textit{Panel (c)}} shows a model giving rise to stronger line profiles with a possible P-Cygni. This spectrum represents the ideal case for the diagnostic diagram methodology, with hints of an asymmetric profile shape and a well-defined single peak at particular inclinations. These types of profiles are typically associated with higher accretion rates, wind mass-loss rates, acceleration exponents and longer acceleration lengths. The observed strength of the line depends significantly on the inclination and wind geometry of the system. For example, when the wind is highly collimated around the rotational axis, observing along this axis will generate a stronger equivalent width emission line than when viewed equatorially. In general, we find more single-peaked profiles at lower inclinations, as expected from basic projection effects. Single-peaked lines also tend to favour a lower wind collimation.

\textbf{\textit{Panel (d)}} shows a model producing a broader, moderately strong line with multiple prominent peaks. These spectra are relatively rare in our grid space, but when they do occur, they are typically associated with line-flux models weaker than the one displayed here. As a result, we can not point to a particular \textsc{sirocco} wind parameter that creates these types of profiles as we did for those in previous panels. Instead, we tend to find such distinctive shapes mainly at the more extreme inclination ranges (very face-on or edge-on) with narrower opening angles (equatorially or jet-like). For this particular model, we have four distinct peaks. These peaks are apparent on the 20° inclined spectrum and less evident at the other inclinations. However, the peaks become apparent on all spectra when measured in normalised flux. This setting is available with our \href{https://austenwallis.pyscriptapps.com/h-alpha-grid-inspector/latest/}{web-based interactive tool}. The wavelength shift of these peaks is related to the projected outflow velocities towards the observer in regions of the model with higher emissivities: Emissivities here essentially mean regions of the wind with higher total emitted flux, which is dependent on several outflow conditions such as higher densities, closer proximity to continuum source, higher optical depth and an optimal ionization balance. Taking the 20° sightline with four peaks, for example, we find two high and distinct emissivity regions, which is possible due to the finite size of the disc. One might, therefore, expect these two spots to be rotationally dominated regions, but this is not the case here. The inner pair of peaks, with a velocity separation of $\approx250\,km\,s^{-1}$, is \textit{indeed} a high emission spot linked to the Keplerian velocity at the outer edge of the accretion disc. However, the outer pair of peaks, with a velocity separation of $\approx650\,km\,s^{-1}$, is emission linked to the poloidal streamline velocities of the outflow at the other high emissivity spot. In essence, forming a bright `equatorial ring-like' effect that creates another double-peaked part of the line \citep{gill_emission-line_1999}. All in all, no matter how interesting these profiles are, fitting a Gaussian to these emission lines is challenging, and our observational benchmark spectra do not display such multi-peaked profiles. We therefore do not include profiles with more than two peaks in our analysis below. 

\subsection{Filtering Spectra into Quality-Tiered Subsamples}
\label{sec:Filtering}

As illustrated by Fig. \ref{fig:sirocco_line_profiles}'s \textit{(a)} and \textit{(d) panels}, our grid of 729 CV models generates some profiles that fall outside the observational parameter space of interest or which violate key assumptions built into the excess-diagnostic method. In order to deal with this, we define four samples based on our grid:

\begin{enumerate}
    \item {\bf FULL}: This sample contains all line profiles from our \textsc{sirocco} simulations with which we can calculate FWHM, EW and/or excess EW quantities.    
    \item {\bf BRONZE}: This sample, a subset of full and the complement set to Silver, shows line profiles with minimal line flux or displays at least three line peaks, placing them outside Silver's selection criteria.
    \item {\bf SILVER}: This sample, a subset of full and complement set to Bronze, contains only single or double-peaked line profiles from which we can derive meaningful EW excess values, \textit{but do not} satisfy the Gold constraints. 
    \item {\bf GOLD}: This sample, a subset of Silver, contains only line profiles that additionally satisfy observationally-based constraints based on the overall FWHM and EW.
\end{enumerate}

Note that membership in a particular sample is inclination-dependent. For example, a model that produces realistic profiles at $i = 60^\circ$ but not at $i = 45^\circ$ may be in the Gold sample for the former inclination, but not for the latter.

\subsubsection{The SILVER Sample}
\label{sec:filter sample}

To define the Silver sample, we employ three filters. These filters remove approximately 50 per cent of the line profiles across the model grid, with the exact value depending on inclination. They are defined as follows:

\begin{enumerate}
    \item \textbf{Removing spectra with no line features} -- Not all spectra within our grid space produce outflows `strong' enough to show any $\mathrm{H\,\alpha}$ spectral line. These spectra are essentially pure continuum emission, are observationally uninteresting and only produce excesses of 0 with noise as the error bar when passed through the excess EW calculation. No observational system will ever be diagnosed as outflowing from a completely featureless spectrum. We therefore discard any spectrum whose strongest emission peak lies less than 1 per cent above the fitted continuum (i.e.\ \(\max[F(\lambda)/C(\lambda) - 1] \le 0.01\)). The number of spectra flagged within this category, depending on the inclination, is usually around 280 to 350 out of 729.

\begin{figure*}
    \centering
    \includegraphics[width=1.0\linewidth]{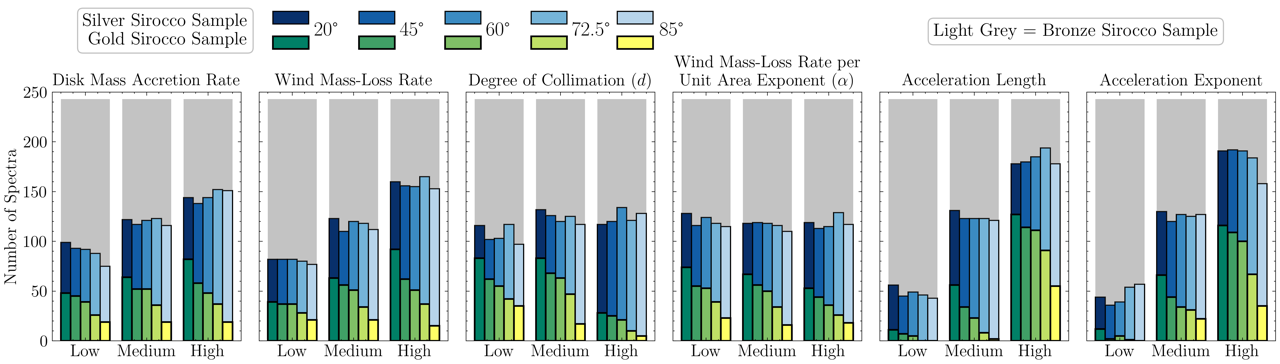}
    \caption{Frequency of \textsc{sirocco} model parameter values in the Gold (Green), Silver (Blue), and Bronze (Grey) samples. Each panel highlights one of the six input parameters. Within each panel, the three x-axis categories (Low, Medium and High) correspond to the parameter's three numeric grid values. Table \ref{tab:combination table} presents the parameter numeric values for each category. Increasing bar lightness corresponds to higher inclined systems from 20° to 85°.}
    \label{fig:combination_bars}
\end{figure*}
    
    \item \textbf{Removing spectra with more than two peaks} -- The excess diagnostic method relies on fitting a Gaussian to the spectral line. However, if the spectral line has multiple peaks, accurately describing this flux distribution with a Gaussian becomes impossible. Fig. \ref{fig:sirocco_line_profiles} shows a variety of $\mathrm{H\,\alpha}$ profiles. \textit{Panel (d)} shows a more unusual triple-peaked emission line. If we fit a Gaussian to the profile, the Gaussian will be biased to one side of the emission line peak, leading to an unintended bias within our excess calculation. Hence, we automatically remove profiles with three or more peaks using the Scipy signal's \textsc{find peaks} Python function. We configure the peak finder only to detect peaks that exceed 30 per cent of the maximum flux value (height). We require a minimum separation of 10 wavelength bins (distance) to ensure neighbouring peaks remain distinct. Additionally, we assess peak prominence, which measures how much a peak stands out from the baseline signal and set this threshold to 10 per cent of the maximum flux. Through inspection, we find these conditions adequately flag and filter lines with three or more peaks that a researcher would likely highlight. However, we do not filter double-peaked spectra at this stage for reasons stated in the next filtering method. The number of spectra flagged within this category, depending on the inclination, is usually around 340 to 390 out of 729.
    
    \item \textbf{Removing poor fits and unrealistic excesses} -- We do not automatically remove double-peaked spectra because of their significant variety and more frequent observation. \textit{Panel (b)} in Fig. \ref{fig:sirocco_line_profiles} provides a clear example of a double-peaked profile that should remain in our analysis. This profile shows a distinct P-Cygni line at particular inclinations with a narrow, equal-height double-peaked top. This spectrum indicates evidence of an outflow that the excess diagnostic method aims to detect. Furthermore, the narrowness of the double peak means the optimal Gaussian fit usually remains centred on the line’s core. However, not all double peaks are suitable. As a result, we measure the pointwise fractional root mean square ($FRMS$) and root mean square ($RMS$) of the residuals following
    \begin{equation*}
        FRMS =\sqrt{{\frac{1}{N}\sum^{N}_{i=1}\frac{(d_i - f_i)}{c_i}^2}}, RMS = \frac{\sqrt{\frac{1}{N}\sum^{N}_{i=1}(d_i-f_i)^2}}{\overline{c}.}
    \end{equation*} 
    where $d$ is the spectral data, $f$ is the Gaussian fit and $c$ is the continuum fit.
    Slightly different to the norm, we divide by continuum values to standardised comparison between spectra. We remove spectra with extreme residuals greater than 0.5. These error values are not due to an asymmetry but rather from poor fits or wide, strong bimodal distributions. Additionally, we remove any spectra whose excess EW errors are greater than $\mathrm{0.5\,\mathring{A}}$. These data points of high error show that the fit is highly dependent on the continuum noise and will again bias the analysis. The number of spectra flagged within this category, depending on the inclination and masking profile, is usually on the order of 1 to 10 out of 729.
    
\end{enumerate}

The frequency of particular \textsc{sirocco} parameter values associated with the Silver sample is highlighted by the blue bars in Fig. \ref{fig:combination_bars}. We find that longer acceleration lengths and higher acceleration exponents are critical in generating significant $\mathrm{H\,\alpha}$ emission lines at all inclinations. As expected, higher accretion and wind mass-loss rates also aid in producing detectable emission lines. Naturally, the vice-versa view applies to the Bronze sample spectra, the complement set to Silver, shown in light grey on the figure. 

Additionally, Fig. \ref{fig:combination_bars}'s green bars highlight the frequency of values associated with our gold sample spectra. These are a sub-selection of Silver line profiles that are observationally similar to our benchmarks. We subsequently discuss the gold selection requirements in the following section.

\subsubsection{The GOLD Sample}
\label{sec:FWHMvsEW}

\begin{figure*}
    \centering
    \includegraphics[width=\linewidth]{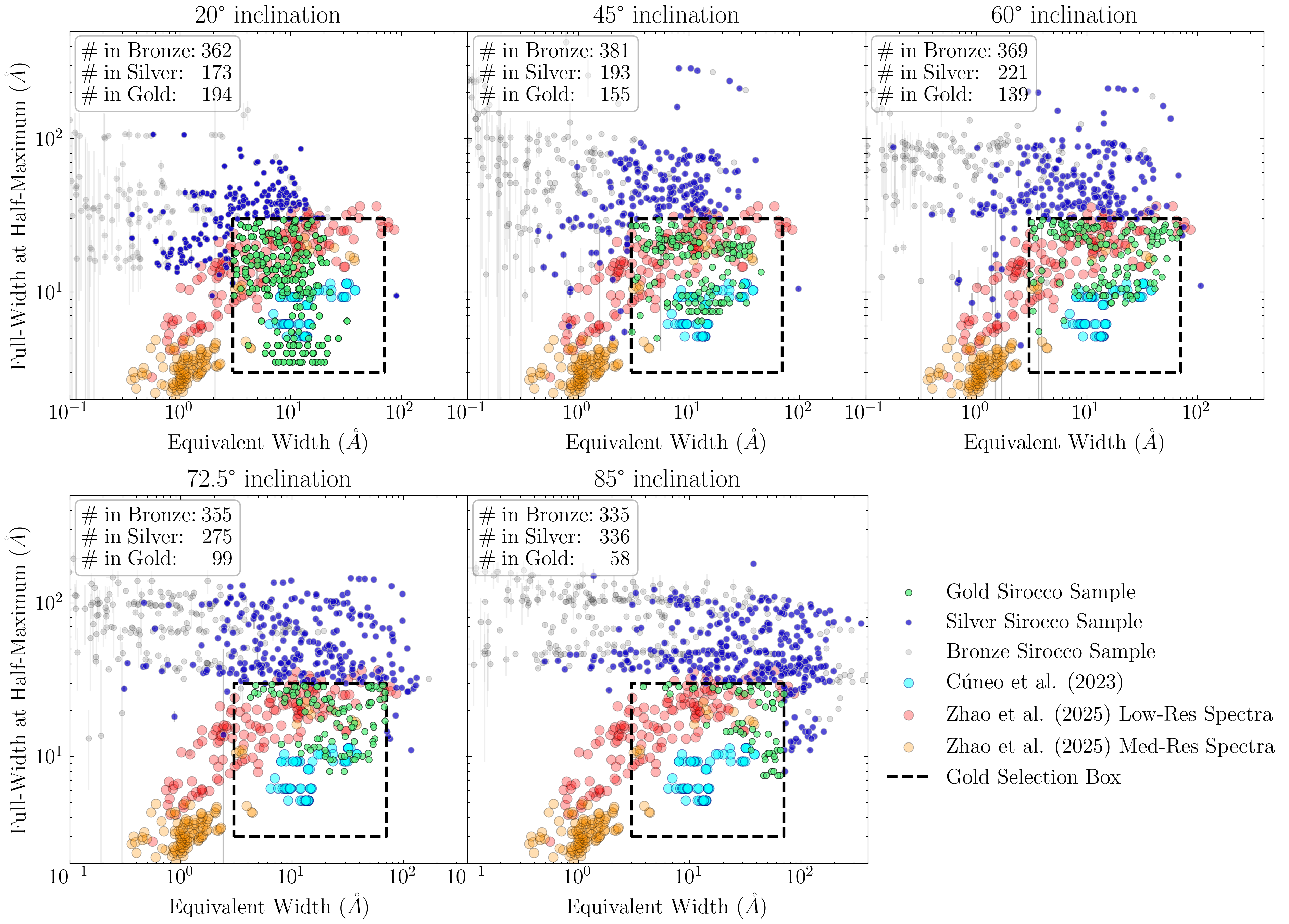}
    \caption{A Full-Width at Half Maximum -- Equivalent Width plot comparing the strength and breadth of \textsc{sirocco} $\mathrm{H\,\alpha}$ line profiles to observed line profiles from \citet{cuneo_unveiling_2023} and \citet{zhao_searching_2025}. Filled circular data points describe line profiles obtained from observing campaigns. Point-like data points describe our \textsc{sirocco} simulations with navy indicating Silver sample spectra and faded light grey showing Bronze sample spectra. Green data points highlight the Silver spectra within the selection box defining the gold subset. We place this selection box (at $\mathrm{3\,\mathring{A} \leq EW \leq  70\,\mathring{A}, 3\,\mathring{A} \leq FWHM\leq 30\,\mathring{A}}$) surrounding \citet{cuneo_unveiling_2023}'s data points to restrict our dynamic range of available \textsc{sirocco} spectra during further analysis. This box ensures we compare only similar line profile characteristics that can describe BZ Cam, MV Lyr, V751 Cyg and V425 Cas on the diagnostic diagrams. For each inclination's panel, an inset annotation displays the total number of spectra within a given sample.}
    \label{fig:EWvsFWHM}
\end{figure*}

To define our Gold sample, we first must define `observationally plausible' line shapes. For this, we calculate the EW and FWHM values for our simulated \textsc{sirocco} line profiles the same way we did for the observational benchmark spectra described in Section \ref{sec:Benchmark}. We then compare the simulated and observed profiles on an FWHM--EW diagram, shown by Fig. \ref{fig:EWvsFWHM}.

When looking at this figure, the first point worth noting is the noticeable systematic differences between the two observational samples. More specifically, \citet{cuneo_unveiling_2023}'s dynamic range in EW and FWHM is considerably smaller. Equally, for a given EW, the FWHMs are smaller by several factors compared to \citet{zhao_searching_2025}'s sample. These differences may be partly due to the different method used by \citet{zhao_searching_2025} to estimate EW and FWHM. However, the main reason is likely to be selection effects. As noted in Section \ref{sec:Benchmark}, the \citet{zhao_searching_2025} sample includes all kinds of CVs. By contrast, the \citet{cuneo_unveiling_2023} sample consists of only 4 high-luminosity nova-like systems, which are relatively rare amongst the CV population.

The second point to note is that the \textsc{sirocco} simulated spectra can fully overlay the FWHM vs EW parameter space defined by the \citet{cuneo_unveiling_2023}, but not the space defined by the full CV sample from \citet{zhao_searching_2025}. This fact is not too surprising, since the parameter space covered by our grid -- in particular $\dot{\mathrm{M}}_{\mathrm{disc}}$ and $\dot{\mathrm{M}}_{\mathrm{wind}}$ -- is designed only to represent high-state CVs (non-magnetic NLs and DN in outbursts). 

We select our observationally plausible spectra by drawing a selection box on Fig. \ref{fig:EWvsFWHM} roughly centred on the \citet{cuneo_unveiling_2023} sample, but with generous margins. We allow a quite liberal margin towards larger FWHM values, so that our box also encompasses most of the \citet{zhao_searching_2025} sample within our adopted EW range. Any \textsc{sirocco} spectra contained within this box form our Gold sample. The EW and FWHM ranges defining the boundaries of our Gold selection box are $\mathrm{3-70\,\mathring{A}}$ and $\mathrm{3-30\,\mathring{A}}$, respectively.

\section{Results}
\label{sec:results}

\begin{figure*}
    \centering
    \includegraphics[width=\textwidth]{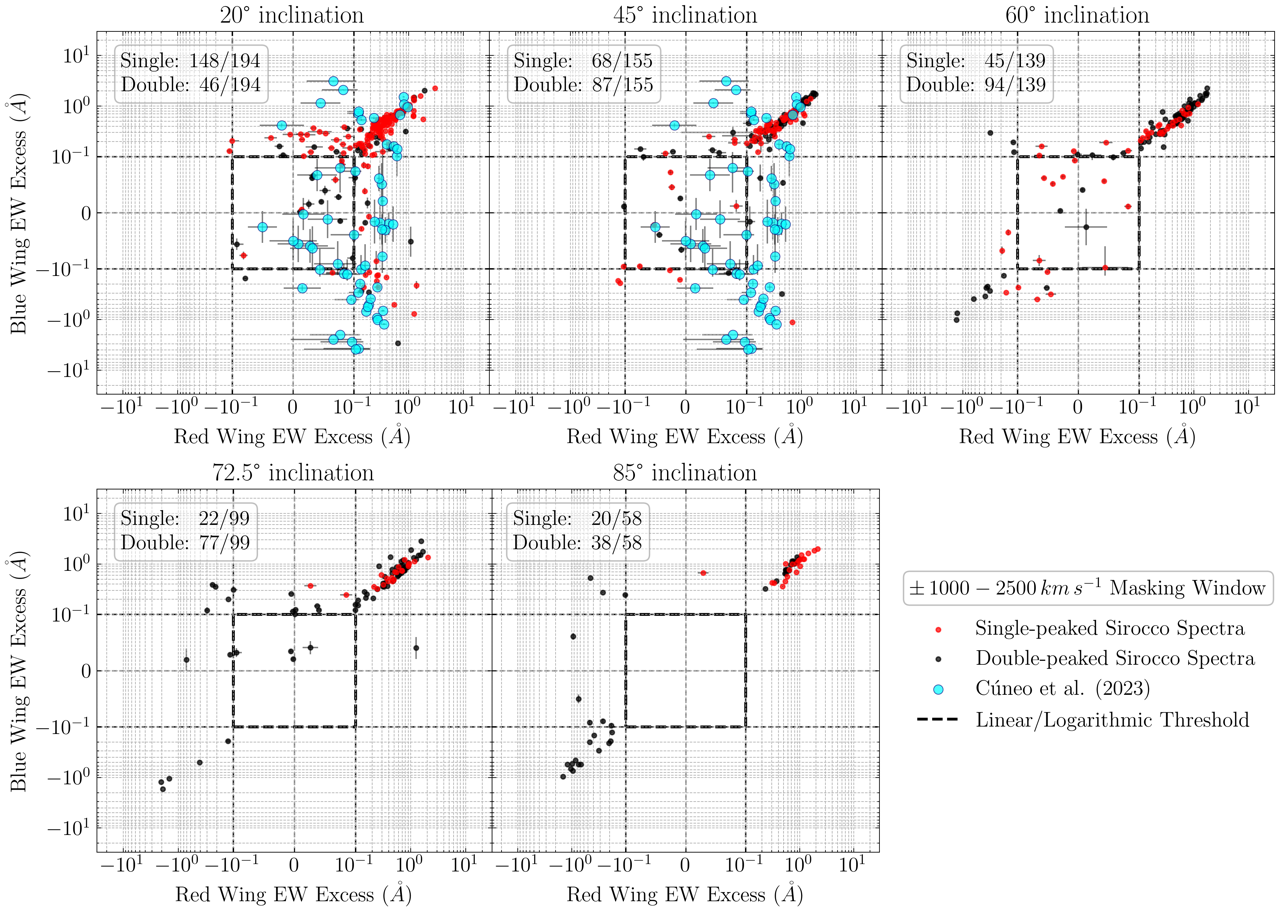}
    \caption{Excess diagnostic diagrams for Gold \textsc{sirocco} CV systems viewed at a 20°, 45°, 60°, 72.5° and 85° inclination. Red data points indicate \textsc{sirocco} spectra that exhibit only one prominent peak within the $\mathrm{H\,\alpha}$ emission line. Black data points indicate \textsc{sirocco} spectra with two prominent peaks. The masking profile is set to a radial velocity at $\pm\,\mathrm{1000-2500\,km\,s^{-1}}$ which corresponds to wavelengths $\pm\,\mathrm{22-55\,\mathring{A}}$ from the rest wavelength of $\mathrm{H\,\alpha}$ ($\mathrm{6562.819\,\mathring{A}}$). The linear-logarithmic threshold for the symmetric logarithmic scale plot, shown as dashed black lines, is set at $\mathrm{0.1\,\mathring{A}}$.}
    \label{fig:gold_diag}
\end{figure*}

Fig. \ref{fig:gold_diag} shows the excess EW diagnostic diagrams for the Gold sample using a $\pm\,\mathrm{1000-2500 \,km\,s^{-1}}$ masking window. Each panels correspond to distinct inclinations at $\mathrm{20^{\circ}}$, $\mathrm{45^{\circ}}$, $\mathrm{60^{\circ}}$, $\mathrm{72.5^{\circ}}$ and $\mathrm{85^{\circ}}$. We also plot the \citet{cuneo_unveiling_2023} sample excesses, but only on the $\mathrm{20^{\circ}}$ and  $\mathrm{45^{\circ}}$ panels since all these sources are thought to be viewed at lower inclinations. Fig. \ref{fig:appendix_diag_plots_22_55} shows the same diagnostic diagrams but also includes the larger Silver sample.

We have chosen to present our diagnostic diagrams on a symmetric logarithmic scale (symlog) plot since many data points cluster at similar small values of EW and FWHM. However, the overall dynamic range is still quite extensive. Briefly, in a symlog plot, the scale varies linearly within a particular range surrounding the origin, but logarithmically outside that range. For our excess EW diagrams, we adopt $\mathrm{\pm\,0.1\,\mathring{A}}$ as the threshold for this linear region. In Fig. \ref{fig:gold_diag}, a black dashed line outlines this boundary. One crucial subtlety of the symlog representation is that areas directly to the north, east, south and west of the central square region have scales that are linear in one axis but logarithmic in the other. 

Overall, Fig. \ref{fig:gold_diag} shows that our wind-formed line profiles shift toward the symmetric $\mathrm{y=x}$ diagonal with increasing inclination. At the same time, double-peaked profiles become more prevalent and the characteristic size of the EW excesses also increases. For example, at $\mathrm{85^{\circ}}$, there are no Gold spectra with both blue and red excess below $0.1\,\mathring{A}$ inside the linear region. In tandem, the largest excess data points become more extreme. This increase is mainly driven by the increasing peak-to-peak separation of the high-inclination double-peaked profiles. These peaks are poorly fit by a single Gaussian and their line cores begin to bleed into the excess masking region. 

Comparing the excess EWs of our Gold sample to the observed values, we find that the models occupy broadly the same region of the diagnostic diagram as the data. Most importantly, at the low inclinations relevant to the \citet{cuneo_unveiling_2023} sample, there is a clear preference for positive red wing excesses. Equally, at these inclinations, the majority of model spectra are single-peaked, also in line with the observed spectra (cf. Fig. \ref {fig:cuneo_line_examples}). On the other hand, a similar majority of model spectra lie near the positive $\mathrm{y=x}$ diagonal. Meanwhile, the observed spectra are much more randomly scattered across the positive red excess region of the diagram. 

EW excesses near this 1:1 diagonal can be produced simply by symmetric line profiles with non-Gaussian wings. This certainly includes disc-formed emission lines, for which the shape of the line wings directly reflects the radial surface brightness profile across the disc \citep{smak_emission_1981, horne_emission_1986}. As a result, only {\em off-diagonal} excesses might potentially indicate disc wind signatures. In this sense, the observed line profiles in the \citet{cuneo_unveiling_2023} data set actually look {\em more} `windy' than those associated with the `average' \textsc{sirocco} wind-formed features. In other words, based on their positioning in the diagnostic diagram, only a relatively small fraction of our disc wind models produce excess EWs that would have been traditionally interpreted as disc wind signatures.

The models also struggle to produce the largest absolute blue excesses associated with the system, BZ Cam. These excesses lie along the $\mathrm{0.1\,\mathring{A}}$ red wing threshold line with blue wing excess emission greater or less than $\mathrm{1\,\mathring{A}}$. As discussed by \citet{matthews_disc_2023} in the context of QSOs, disc winds {\em can} produce asymmetric and systematically blue-shifted emission lines. Such lines would be expected to lie in the strong blue excess region of the diagnostic diagram. Physically, this occurs when the receding half of the outflow is blocked from view. In compact binaries, we might expect this to occur primarily in low inclination systems, where we can only see the `top' half of the disc and wind. The optically thick disc obscures our view of the other side. A few models in our grid do seem to show this effect, but not as strongly as suggested by the data. However, we are not surprised that our \textsc{sirocco} data points do not \textit{exactly} match and scatter around the \citet{cuneo_discovery_2020} sample, given the generic nature of a systematic grid search. If anything, we expect that. For these few BZ Cam data points, we can conclude that either these winds are more extreme than our standard expectation, or another different emission line component contributes significantly to the observed line profiles.

\section{Discussion}
\label{sec:discussion}

\subsection{Excess EW Diagram Reliability for Outflow Diagnostics}

The observed line profiles of wind-driving high-state CVs place them in a region of the diagnostic diagram previously suggested as an area indicating outflows. This outflow region is characterised by relatively large, preferentially red-wing EW excesses that lie off the diagonal of the diagram. So, do our line profiles produced by \textsc{sirocco}'s disc wind models support this idea? 

The two main conclusions we can draw from the results in Section \ref{sec:results} are that our disc wind models {\em can} occupy this same region of the diagram, but most do not. They instead lie near the diagonal as expected for symmetric profiles with non-Gaussian wings. Taken at face value, this could indicate that the observed excesses {\em are} indicative of outflow but that these outflows must have fairly specific physical properties.

However, these two conclusions should be taken with a large grain of salt. As we see in the next section, the excess EW diagnostic is somewhat sensitive to the exact definition of the masking window used to remove the line core in the excess EW calculation. Most of our profiles lie on the diagonal because of the adopted masking window choice at $\pm\,\mathrm{1000-2500 \,km\,s^{-1}}$. In many cases, the line profiles have sizeable FWHMs, so their cores begin to bleed into the masking windows that calculate the excesses. The effect is that excesses from the strong core emission can dominate over any slight wing asymmetry, pushing the data points towards the diagonal.

Additionally, we should remember that line formation in a disc wind is not the only way to produce asymmetric line wings. Except for apparent blue-shifted absorption, no asymmetry can be considered a `smoking gun' signature of an outflow. For example, any asymmetric structure in the accretion flow -- perhaps associated with the `hot spot' where the accretion stream impacts the disc edge -- could give rise to excess EW in the blue or red wing of an otherwise disc-dominated line. Similarly, an eccentric accretion disc will also produce asymmetric line wings. Therefore, we caution against interpreting weak asymmetries as clear {\em evidence} for an outflow. Without an actual P-Cygni profile or apparent blue-shifted absorption, the closest to a wind {\rm signature} combines a blue deficit with a red excess.

Regardless, in Section \ref{sec:FWHM_masking}, we will suggest a revised definition based on the FWHM of a line which does not suffer from this core contamination. The Gold sample spectra in the revised diagnostic diagrams still cover the same regions as the observations. However, the diagrams' appearance and the set of models overlapping with the observations significantly differ from Fig. \ref{fig:gold_diag}. 

\subsection{Sensitivity to the Adopted Masking Regions}

The user-chosen radial velocity masking choice can drastically impact the data point distribution. Researchers have applied a variety of masking windows across several studies from a small $\mathrm{\pm\,500-1000\,km\,s^{-1}}$ window to up to broad $\mathrm{\pm\,500-4000\,km\,s^{-1}}$ window \citep{panizo-espinar_optical_2021, mata_sanchez_1989_2018, mata_sanchez_ask_2023}. Comparing the broadest radial velocity range studied to the $\pm\,\mathrm{1000-2500 \,km\,s^{-1}}$ velocity window used in the previous section, we find a complete reshuffling of the EW excess data. This change in window comparison can be inspected from Fig. \ref{fig:appendix_diag_plots_22_55} and Fig. \ref{fig:appendix_diag_plots_11_88}. The data points populate more towards the corners of the diagnostic diagrams with a broader window. This broader window includes more of the line's core and more extreme wing discrepancies. As a result, the core of stronger emission lines, particularly with Gaussians fitted to prominent double-peaked spectra, significantly contributes to the total excess EW. This core contribution is usually far beyond any asymmetric contribution from the emission wings we wish to diagnose. Therefore, any conclusions we derived from the previous $\mathrm{\pm\,1000-2500\,km\,s^{-1}}$ masking window will no longer hold in this broader $\mathrm{\pm\,500-4000\,km\,s^{-1}}$ masking scenario. This example utilises a pretty significant change in the masking window. However, this effect occurs even with minor changes in the masking window.

\begin{figure}
    \centering
    \includegraphics[width=1.0\linewidth]{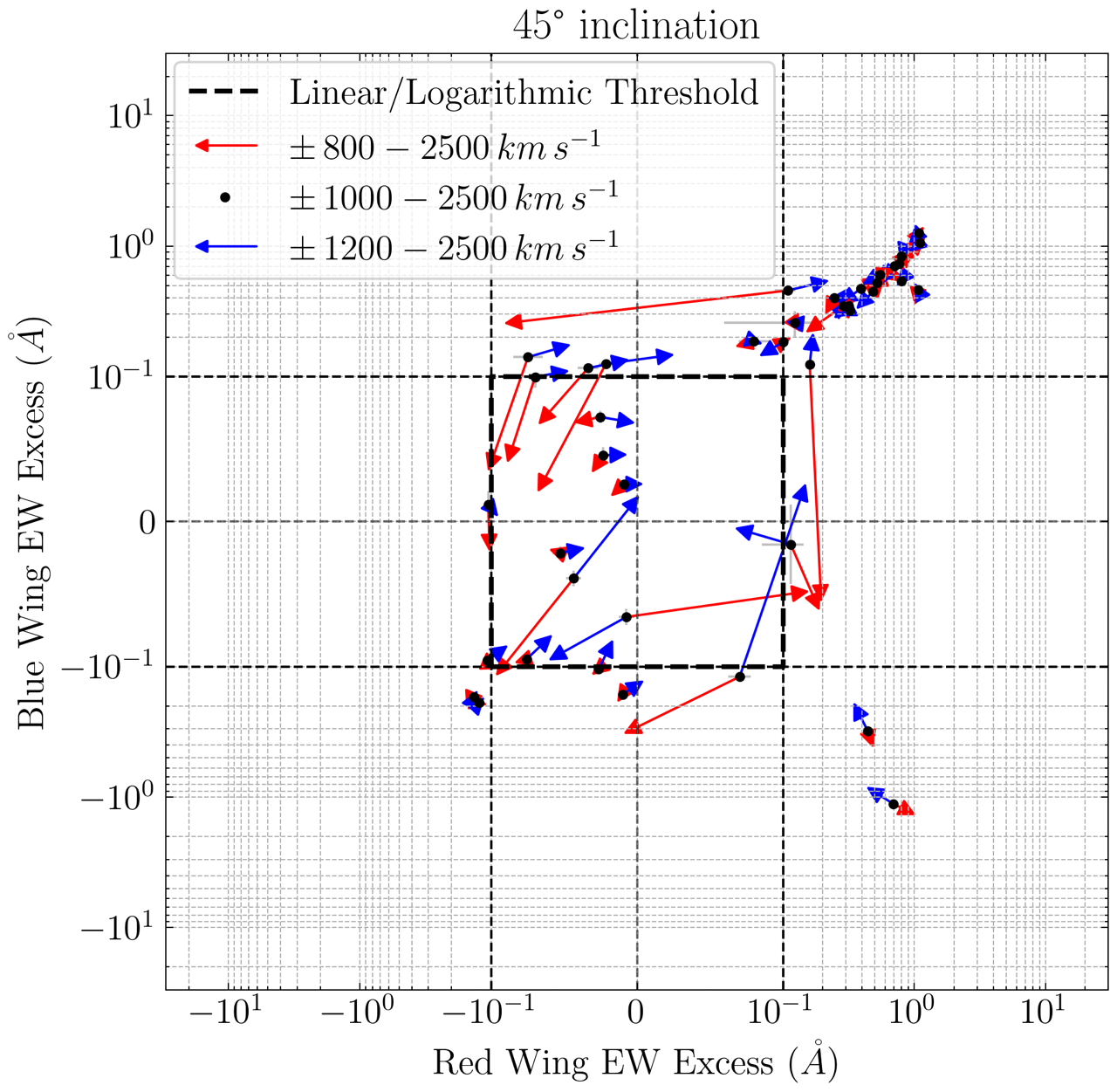}
    \caption{For a representative random subset selection of line profiles, a diagnostic diagram showing the change in excess EWs for a $\mathrm{200\,km\,s^{-1}}$ change in the radial velocity's inner masking window edge. The black data points are calculated from a $\pm\,\mathrm{1000-2500\,km\,s^{-1}}$, matching the diagnostic diagram within this study. The red arrow indicates the change in the data points position, given the $\mathrm{200\,km\,s^{-1}}$ radial velocity boundary shift towards the core of the spectral line, increasing the window width. The blue arrow represents the change in position when the inner bound shifts $\mathrm{200\,km\,s^{-1}}$ away from the line core, decreasing the window width.}
    \label{fig:masking_effect}
\end{figure}

To display this issue more prominently, Fig. \ref{fig:masking_effect} shows a slight change in the masking profile for a small random selection of data points. In this figure, we plot 40 \textsc{sirocco} spectra and measure the EW excesses with three different masking windows. These masks change by $\mathrm{\pm\,4\,\mathring{A}}$ near the line's core, corresponding to an approximate $\mathrm{\pm\,200\,km\,s^{-1}}$ change in radial velocity, following $\Delta\lambda = \frac{v_r\lambda_0}{c}$. The outer window boundary remains the same. We plot the $\pm\,\mathrm{1000-2500\,km\,s^{-1}}$ masking window as a data point, with arrows pointing to the data point's new position when slightly adjusting the inner boundary. As shown, several data points remain largely static within the plotting regime, which is good. However, a sizeable minority of spectra move drastically across the diagram, with 11 of the 155 data points switching plotting quadrants entirely. This pronounced change in position for such a slight change in the window is undesirable, especially given that astronomical effects, such as redshift, can alter the rest wavelength position of a spectral line. We also find similar shifting effects when sliding a fixed width masking window from $\pm\,\mathrm{900-2400\,km\,s^{-1}}$ to $\pm\,\mathrm{1100-2600\,km\,s^{-1}}$ and with changes shifting both inner and outer edges by $\mathrm{\pm\,100\,km\,s^{-1}}$, squishing and stretching the window. This observation highlights the need for an adaptive masking window approach that changes with the line's intrinsic core width, thereby minimizing artificial shifts by the fixed window. 

With this plot, the reader should remember the symlog scaling critically. A natural effect of this scale is that arrows will tend to increase in length the closer they are to the origin. Equally, tiny arrows far from the origin can represent significant value changes. This effect, without consideration, will allow the reader to over-emphasize the importance of large arrows near the centre and under-emphasize smaller arrows on the outer parameter. We find there tends to be a consistent and noticeable change in excess values for a slight shift in masking profile across the \textsc{sirocco} grid space. 

Given that we choose a subset of 40 \textsc{sirocco} spectra, a question is likely to be proposed regarding how tailor-made Fig. \ref{fig:masking_effect} is to illustrate this effect. We attempt to display a representative example of the population that avoids overlapping arrows. We break the data points into their respective four plotting quadrant groups. Then, we plot all data points in the north-west, south-west and south-east quadrants. Then, we randomly choose 20 out of the 134 data points from the north-east quadrant. This region contains the largest concentration of our data points. We repeat this random selection process to find and display a reasonable-looking diagram. However, this figure could be shown entirely with a subset of data points containing large criss-crossing arrows across the diagram. Equally, we could show a wholly stationary subset. 

Overall, we find fixed mask boundaries unsuitable for comparing astronomical systems. Researchers could argue for using fixed bounds when measuring an individual system's change in outflow properties, such as in \citet{mata_sanchez_1989_2018}'s transient scenario of V404 Cygni. However, we likely require a different non-user-chosen or dynamic masking window methodology when measuring across different systems. Hence, as a natural first step, we implement the masking window as a function of the line's FWHM.

\subsection{Reliability Improvements with FWHM-Based Masking}
\label{sec:FWHM_masking}

Fig. \ref{fig:sirocco_line_profiles} and Fig. \ref{fig:EWvsFWHM} illustrate that our \textsc{sirocco} models have a notable variety in line profiles and a wide range of full-width at half maximums. Intrinsically, defining a masking window at pre-determined radial velocities does not make sense when attempting to systematically detect outflowing systems through asymmetric wings, especially if incoming data are for a survey-like campaign with thousands of spectra. We need a more reliable method of selecting the masking window to solely bind the spectral line wings. Instead, a natural choice is to implement a dynamic masking window as a function of the spectral line's FWHM. Hence, we implement this idea and analyse the impact of this masking choice on the diagnostic diagram analysis. 

\begin{figure*}
    \centering
    \includegraphics[width=\linewidth]{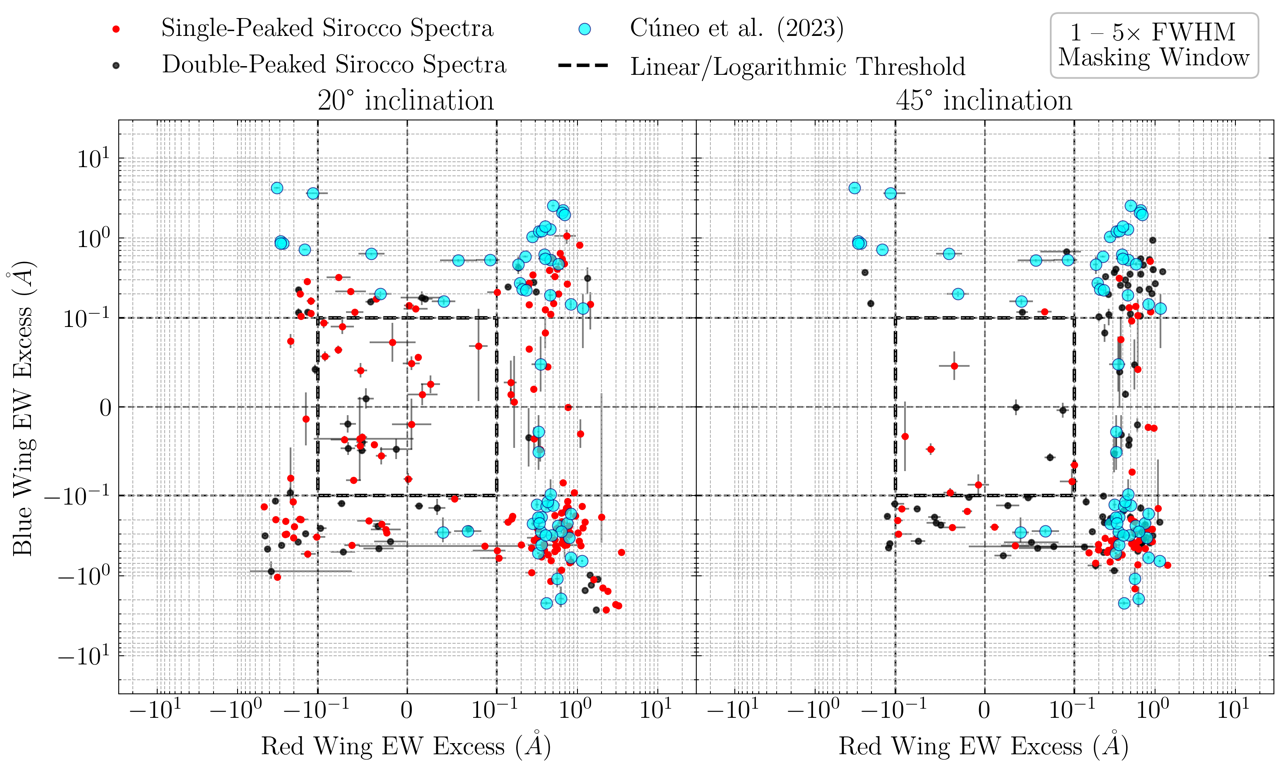}
    \caption{FWHM masking windowed excess diagnostic diagrams for Gold \textsc{sirocco} CV systems viewed at a 20° and 45° inclination. Red data points indicate \textsc{sirocco} spectra that exhibit only one prominent peak within the $\mathrm{H\,\alpha}$ emission line. Black data points indicate \textsc{sirocco} spectra with two prominent peaks. The masking profile bounds are dynamically set at $\mathrm{1.0\times}$ and $\mathrm{5.0\times}$ of a line profile's respective FWHM from the rest wavelength of $\mathrm{H\,\alpha}$ ($\mathrm{6562.819\,\mathring{A}}$). The linear-logarithmic threshold for the symmetric logarithmic scale plot, shown as dashed black lines, is set at $\mathrm{0.1\,\mathring{A}}$.}
    \label{fig:Figure_FWHM_diag}
\end{figure*}

We create the masking window as a function of the recorded FWHM values following the method described in Section \ref{sec:FWHMvsEW}. We set the inner boundary at $\mathrm{1\times FWHM}$ and the outer boundary at $\mathrm{5\times FWHM}$. Arguably, this window function is still user-chosen as we set the boundaries at any multiple. However, our choice guarantees we avoid substantial contamination from the line core for all spectra and include the same ratio of the line's wings for all excess measurements. This choice allows for a more consistent analysis of the same line profile characteristics across systems without changing the diagnostic masking methodology and reduces the overall sensitivity of excess measurements to the window choice. Similarly, but to a lesser extent than the fixed mask scenario, the excess measurements are more sensitive to the inner boundary choice at $\mathrm{1\times FWHM}$ than the outer boundary at $\mathrm{5\times FWHM}$ due to the bounds' closer proximity to the line's core. 

We implement the FWHM masking window methodology for the diagnostic diagrams in Fig. \ref{fig:Figure_FWHM_diag}. Here, we only show the 20° and 45° inclinations for easier data point comparison. However, all inclinations can be seen in Fig. \ref{fig:appendix_diag_plots_fwhm}. We find that the reliability improves when predicting line profile characteristics given a data point's position within this new diagnostic framework. Data points in the south-east quadrant, or the expected P-Cygni region illustrated by our cartoon in Fig. \ref{fig:diag_example}, are predominantly P-Cygni profiles. This logic applies across the diagnostic diagram. One region of particular interest is the extreme \citet{cuneo_unveiling_2023} data points in the north-west quadrant. Typically, this region, defined by enhanced blue wings and suppressed red wings, is an unusual characteristic for outflowing spectra to populate. However, these spectra do indeed reflect large blue and negative red excesses upon their line profiles. The original fixed masking window method does not convey this asymmetric fact. We still note how our \textsc{sirocco} spectra do not populate this extreme region. This result supports our earlier thought that wind and disc emission can not entirely describe all \citet{cuneo_unveiling_2023} spectra. 

Another advantage of the FWHM masking window is the greater spread of data points across the diagram. We no longer focus the data points along the $\mathrm{y=x}$ symmetric line profile axis. Line core contamination is the primary driver of spectral excesses towards the diagonal. Hence, the reduction in this effect proves this methodological improvement towards the sole measurement of the asymmetric wings. Nevertheless, the masking window is still imperfect, as highlighted by higher inclinations. Stronger double-peaked profiles dominate these high inclinations and populate the south-west suppressed wings quadrant. However, this population is not, in essence, a fault of the masking window, but rather of the Gaussian fitting process to a double-peaked emission line. We should raise questions about whether strong double-peaked profiles should qualify for diagnostic diagram analysis. 

Overall, this FWHM-based methodology does improve diagnostic diagram reliability for single-peaked line profile characteristics and lower inclined systems.

\subsection{Wind -- EW Scaling Relation as Another Outflow Diagnostic}

\label{sec:EM and EW}

\begin{figure}
    \centering
    \includegraphics[width=\linewidth]{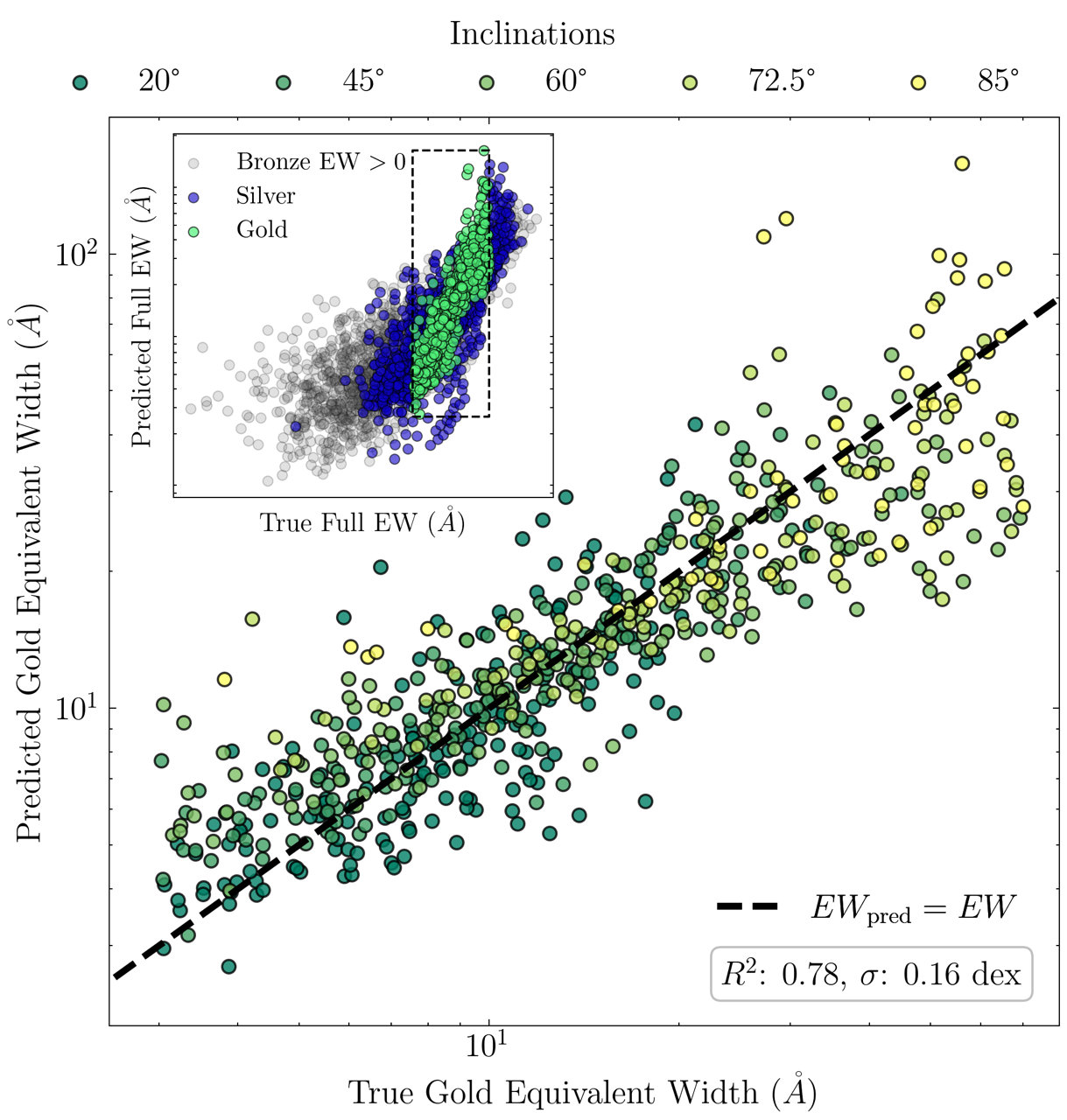}
    \caption{Predictive performance of an approximate calibrated scaling power-law relation for $\mathrm{H\,\alpha}$ line's EW given a particular combination of \textsc{sirocco} parameters. The relation is solely fitted to our Gold sample at all inclinations, which is shown in hues of green. The black dashed $y=x$ line highlights an ideal 1:1 scenario where predicted EW equals the true EW. The inset plot shows the degrading performance when the same relation is applied to the Silver (navy) and Bronze (grey) samples. The dashed black box outlines the Gold sample region. The main panel is a zoomed-in view of that boxed region.}
    \label{fig:Scaling Law}
\end{figure}

The excess EW diagnostic diagram is designed to allow the detection of weak outflow signatures in optical emission lines. The underlying idea here is that the effect of the outflow may be limited to distorting a line profile formed elsewhere in the system, for example, the accretion disc. However, in our models, {\em the entire line} is formed in the outflow.\footnote{Strictly speaking, the distinction between `the accretion disc' and `the outflow' is somewhat artificial. A better argument is to think in terms of an {\em accretion disc atmosphere} that extends from the hydrostatic disc mid-plane all the way to the supersonic outflow.}
In this context, a natural question arises: can the {\em global} properties of an emission line, such as the overall line strength, serve as outflow diagnostics? More specifically, does the emission line EW depend on the physical properties of the outflow like mass-loss rate, collimation and velocity?

For our Gold sample of model line profiles, a scaling relation based on a simple power-law {\em Ansatz} is sufficient to predict the $\mathrm{H\,\alpha}$ EWs to within 0.17 dex RMS (corresponding to $\simeq$50\%) across all inclinations simultaneously. The fitted scaling relation is illustrated in Fig. \ref{fig:Scaling Law} and given by 
\begin{equation}
\begin{split}
\label{eq:scaling}
\mathrm{EW \simeq 5\,\mathring{A}} ~ & \left[\frac{\dot{M}_{acc}}{\mathrm{10^{-8}~M_{\odot}~yr^{-1}}}\right]^{-0.12}
~\left[\frac{\dot{M}_{wind}}{\mathrm{10^{-9}~M_{\odot}~yr^{-1}}}\right]^{0.31}\\
& \left[\frac{d}{\mathrm{5.5~R_{WD}}}\right]^{0.11}
~\left[\frac{R_s}{\mathrm{10~R_{WD}}}\right]^{0.16}\\
& \left[10^{0.31(\alpha-0.25)}\right]
~ \left[10^{0.1(\beta-1.5)}\right]
~ \left[\frac{\cos{i}}{\cos{60^\degree}}\right]^{-0.56}.
\end{split}
\end{equation}

This relation provides observers with a simple way to assess whether -- and what kind of -- accretion disc wind might produce the $\mathrm{H\,\alpha}$ lines in any particular system. However, we stress that there are numerous caveats when using this expression. For example, the fit is only relevant to the KWD disc wind parametrization and is not equally good across the entire parameter range. Additionally, the inset in Fig. \ref{fig:Scaling Law} illustrates that our relation does not continue providing a good prediction for the Silver and Bronze samples as data points begin to diverge away from the ideal line. 

In Appendix \ref{sec:CoG Appendix}, we show that a (slightly) more physically motivated approach based on the volumetric emission measure of the outflow is capable of predicting EWs with reasonable accuracy across our entire set of models (Gold, Silver and also Bronze). However, the Silver and Bronze models are less relevant for interpreting observations and the resulting fitting formulae for predicting them are considerably more complex. Therefore, the approach outlined in Appendix \ref{sec:CoG Appendix} may be more interesting in principle than useful in practice.

\section{Summary and Conclusions}
\label{sec:conclusion}

The main aim of this work has been to test, refine and develop observational diagnostics for accretion disc winds in cataclysmic variables based on their optical $\mathrm{H\,\alpha}$ emission line profiles. Such diagnostics allow the detection and/or characterisation of these outflows. In order to accomplish this, we have created a grid of synthetic wind-formed line profiles using the \textsc{sirocco} ionization and radiative transfer code. The grid covers a wide range of system and outflow parameters based on a simple parametrized model for biconical disc winds. Our main results are as follows:

\begin{enumerate}
\item[$\bullet$] Disc wind models can produce $\mathrm{H\,\alpha}$ emission line profiles that achieve verisimilitude with observations. In particular, we define a Gold sample of wind models lines with EW and FWHM values that are consistent with observations. Our grid also includes model $\mathrm{H\,\alpha}$ line profiles with clear P-Cygni (blue-shifted absorption) signatures. 
\item[$\bullet$] The line profiles in our Gold sample cover essentially the full region occupied by the excess EW diagnostic diagram proposed by \citet{mata_sanchez_1989_2018} and applied to CVs by \citet{cuneo_unveiling_2023}. However, most synthetic line profiles cluster along the diagonal in this diagram, corresponding to symmetric but non-Gaussian line profile shapes. Such shapes cannot be used as reliable outflow signatures in practice.
\item[$\bullet$] We have identified a significant sensitivity of the excess EW method to the user-defined velocity masking window over which the excess is calculated. Therefore, we propose a refinement to the method in which this window is explicitly related to the FWHM of the line profile. The resulting diagnostic diagrams look distinctly different, with a much higher fraction of sources located in the `blue deficit/red excess' region that represents the most compelling outflow signatures. Consequently, the revised definition of the excess EW is more likely to provide a sensitive and reliable way to detect disc winds via optical spectroscopy.
\item[$\bullet$] In our models, the {\em entire} $\mathrm{H\,\alpha}$ line is formed in the accretion disc wind, usually near the dense base of the outflow. We have therefore constructed a convenient approximate scaling relation for the EWs of wind-formed $\mathrm{H\,\alpha}$ lines in CV as a function of the system and wind parameters that define our grid. This provides a simple way to determine what kind of accretion disc wind (if any) might produce the observed $\mathrm{H\,\alpha}$ line in any given CV.
\end{enumerate}

Ultimately, of course, the goal should be to move beyond the use of `diagnostics' entirely. A more physically motivated and elegant approach would be to directly model observed spectra with the kind of disc wind models utilised in this study. The main challenges for such an approach are (i) the computational time required for a minimum standard simulation ($\approx1\,h$ with 32 CPUs). Naturally, higher resolution or denser wind simulations can take up to a day on HPC clusters; (ii) the large number of system and wind parameters require many simulations. Our \textsc{Sirocco} grid presented in this paper incorporates 6 of the 18+ possible parameters. A complete grid space with the minimum simulation requirements would take 22,000 years; (iii) the difficulty in constructing an appropriate goodness-of-fit metric that allows, in a principled way, for `systematic uncertainties' on the models (e.g. associated with the model parametrization or limitations in the atomic data). In an attempt to overcome these challenges, we are currently developing an emulator-based inference method built on the \textsc{starfish} framework \citep{czekala_constructing_2015}.

\section*{Acknowledgements}

This work was supported by the Science and Technology Facilities Council [studentship project reference 2750006]. The authors acknowledge the use of the IRIDIS High Performance Computing Facility, and associated support services at the University of Southampton, in the completion of this work. We thank the authors of \citet{cuneo_unveiling_2023} for generously providing their observational data for verification and comparison within this study. The authors also thank Amin Mosallanezhad for their aid in the examination of this work's reviewer comments, Nicolas Scepi as part of the \textsc{sirocco} collaboration and Madeleine-Mai Ward for their valuable discussions. James Matthews acknowledges funding from a Royal Society University Research Fellowship (URF$\backslash$R1$\backslash$221062). Partial support for Knox Long's effort on the project was provided by NASA through grant numbers HST-GO-16489 and HST-GO-16659 and from the Space Telescope Science Institute, which is operated by AURA, Inc., under NASA contract NAS 5-26555. The authors acknowledge the use of numpy \citep{harris_array_2020}, scipy \citep{virtanen_scipy_2020}, scikit-learn \citep{pedregosa_scikit-learn_2011} and matplotlib \citep{hunter_matplotlib_2007} in the completion of this work.

\section*{Data Availability}
\label{sec:Data}

The data analysis scripts that produce this work are available to view through the  \href{https://github.com/AustenWallis/CV_Asymmetries_Paper_2025}{CV Asymmetries Paper 2025} GitHub repository. Here, one can also find the simulated \textsc{sirocco} models used for this analysis. A \href{https://austenwallis.pyscriptapps.com/h-alpha-grid-inspector/latest/}{web-based interactive tool} is available to browse all our $\mathrm{H\,\alpha}$ spectral profiles and explore the impact of individual parameters. A comprehensive collection of model and calculated data for this study can be found in the paper's supplementary material. An archived version of the GitHub repository can be found via Zenodo (\href{https://doi.org/10.5281/zenodo.15257396}{DOI:10.5281/zenodo.15257396}). \textsc{sirocco} is a freely available software via \href{https://github.com/sirocco-rt/sirocco}{GitHub} with documentation on \href{https://sirocco-rt.readthedocs.io/en/}{ReadTheDocs}. This paper utilises a pre-release version of \textsc{sirocco}, formally known as \textsc{Python v87f}. \\




\bibliographystyle{mnras}
\bibliography{references} 

\begin{thebibliography}{}
\makeatletter
\relax
\def\mn@urlcharsother{\let\do\@makeother \do\$\do\&\do\#\do\^\do\_\do\%\do\~}
\def\mn@doi{\begingroup\mn@urlcharsother \@ifnextchar [ {\mn@doi@} {\mn@doi@[]}}
\def\mn@doi@[#1]#2{\def\@tempa{#1}\ifx\@tempa\@empty \href {http://dx.doi.org/#2} {doi:#2}\else \href {http://dx.doi.org/#2} {#1}\fi \endgroup}
\def\mn@eprint#1#2{\mn@eprint@#1:#2::\@nil}
\def\mn@eprint@arXiv#1{\href {http://arxiv.org/abs/#1} {{\tt arXiv:#1}}}
\def\mn@eprint@dblp#1{\href {http://dblp.uni-trier.de/rec/bibtex/#1.xml} {dblp:#1}}
\def\mn@eprint@#1:#2:#3:#4\@nil{\def\@tempa {#1}\def\@tempb {#2}\def\@tempc {#3}\ifx \@tempc \@empty \let \@tempc \@tempb \let \@tempb \@tempa \fi \ifx \@tempb \@empty \def\@tempb {arXiv}\fi \@ifundefined {mn@eprint@\@tempb}{\@tempb:\@tempc}{\expandafter \expandafter \csname mn@eprint@\@tempb\endcsname \expandafter{\@tempc}}}

\bibitem[\protect\citeauthoryear{Beristain, Edwards  \& Kwan}{Beristain et~al.}{2001}]{beristain_helium_2001}
Beristain G.,  Edwards S.,   Kwan J.,  2001, \mn@doi [The Astrophysical Journal] {10.1086/320233}, 551, 1037

\bibitem[\protect\citeauthoryear{Boroson \& Green}{Boroson \& Green}{1992}]{boroson_emission-line_1992}
Boroson T.~A.,  Green R.~F.,  1992, \mn@doi [The Astrophysical Journal Supplement Series] {10.1086/191661}, 80, 109

\bibitem[\protect\citeauthoryear{Cordova \& Mason}{Cordova \& Mason}{1982}]{cordova_high-velocity_1982}
Cordova F.~A.,  Mason K.~O.,  1982, \mn@doi [The Astrophysical Journal] {10.1086/160291}, 260, 716

\bibitem[\protect\citeauthoryear{Czekala, Andrews, Mandel, Hogg  \& Green}{Czekala et~al.}{2015}]{czekala_constructing_2015}
Czekala I.,  Andrews S.~M.,  Mandel K.~S.,  Hogg D.~W.,   Green G.~M.,  2015, \mn@doi [The Astrophysical Journal] {10.1088/0004-637X/812/2/128}, 812, 128

\bibitem[\protect\citeauthoryear{Cúneo et~al.,}{Cúneo et~al.}{2020}]{cuneo_discovery_2020}
Cúneo V.~A.,  et~al., 2020, \mn@doi [Monthly Notices of the Royal Astronomical Society] {10.1093/mnras/staa2241}, 498, 25

\bibitem[\protect\citeauthoryear{Cúneo et~al.,}{Cúneo et~al.}{2023}]{cuneo_unveiling_2023}
Cúneo V.~A.,  et~al., 2023, \mn@doi [Astronomy \& Astrophysics] {10.1051/0004-6361/202347265}, 679, A85

\bibitem[\protect\citeauthoryear{Dennis \& Phillips}{Dennis \& Phillips}{2024}]{dennis_emission_2024}
Dennis B.~R.,  Phillips K. J.~H.,  2024, Emission {Measures} {Demystified}, \mn@doi{10.48550/arXiv.2403.14845}, \url {http://arxiv.org/abs/2403.14845}

\bibitem[\protect\citeauthoryear{Drew}{Drew}{1987}]{drew_inclination_1987}
Drew J.~E.,  1987, \mn@doi [Monthly Notices of the Royal Astronomical Society] {10.1093/mnras/224.3.595}, 224, 595

\bibitem[\protect\citeauthoryear{Edwards, Fischer, Hillenbrand  \& Kwan}{Edwards et~al.}{2006}]{edwards_probing_2006}
Edwards S.,  Fischer W.,  Hillenbrand L.,   Kwan J.,  2006, \mn@doi [The Astrophysical Journal] {10.1086/504832}, 646, 319

\bibitem[\protect\citeauthoryear{Filiz~Ak et~al.,}{Filiz~Ak et~al.}{2014}]{filiz_ak_dependence_2014}
Filiz~Ak N.,  et~al., 2014, \mn@doi [The Astrophysical Journal] {10.1088/0004-637X/791/2/88}, 791, 88

\bibitem[\protect\citeauthoryear{Frank, King  \& Raine}{Frank et~al.}{2002}]{frank_accretion_2002}
Frank J.,  King A.,   Raine D.~J.,  2002, Accretion Power in Astrophysics, p.~398

\bibitem[\protect\citeauthoryear{Froning}{Froning}{2004}]{froning_observations_2004}
Froning C.~S.,  2004, Observations of {Outflows} in {Cataclysmic} {Variables}, \mn@doi{10.48550/arXiv.astro-ph/0410200}, \url {http://arxiv.org/abs/astro-ph/0410200}

\bibitem[\protect\citeauthoryear{Gill \& O'brien}{Gill \& O'brien}{1999}]{gill_emission-line_1999}
Gill C.~D.,  O'brien T.~J.,  1999, \mn@doi [Monthly Notices of the Royal Astronomical Society] {10.1046/j.1365-8711.1999.02681.x}, 307, 677

\bibitem[\protect\citeauthoryear{Greiner, Tovmassian, Di~Stefano, Prestwich, González-Riestra, Szentasko  \& Chavarría}{Greiner et~al.}{1999}]{greiner_transient_1999}
Greiner J.,  Tovmassian G.~H.,  Di~Stefano R.,  Prestwich A.,  González-Riestra R.,  Szentasko L.,   Chavarría C.,  1999, \mn@doi [Astronomy and Astrophysics] {10.48550/arXiv.astro-ph/9812054}, 343, 183

\bibitem[\protect\citeauthoryear{Groot, Rutten  \& van Paradijs}{Groot et~al.}{2004}]{groot_spectrophotometric_2004}
Groot P.~J.,  Rutten R. G.~M.,   van Paradijs J.,  2004, \mn@doi [Astronomy and Astrophysics] {10.1051/0004-6361:20031771}, 417, 283

\bibitem[\protect\citeauthoryear{Hall et~al.,}{Hall et~al.}{2002}]{hall_unusual_2002}
Hall P.~B.,  et~al., 2002, \mn@doi [The Astrophysical Journal Supplement Series] {10.1086/340546}, 141, 267

\bibitem[\protect\citeauthoryear{Harris et~al.,}{Harris et~al.}{2020}]{harris_array_2020}
Harris C.~R.,  et~al., 2020, \mn@doi [Nature] {10.1038/s41586-020-2649-2}, 585, 357

\bibitem[\protect\citeauthoryear{Hartley, Drew, Long, Knigge  \& Proga}{Hartley et~al.}{2002}]{hartley_testing_2002}
Hartley L.~E.,  Drew J.~E.,  Long K.~S.,  Knigge C.,   Proga D.,  2002, \mn@doi [Monthly Notices of the Royal Astronomical Society] {10.1046/j.1365-8711.2002.05277.x}, 332, 127

\bibitem[\protect\citeauthoryear{Heap et~al.,}{Heap et~al.}{1978}]{heap_iue_1978}
Heap S.~R.,  et~al., 1978, \mn@doi [Nature] {10.1038/275385a0}, 275, 385

\bibitem[\protect\citeauthoryear{Holm, Panek  \& Schiffer}{Holm et~al.}{1982}]{holm_ultraviolet_1982}
Holm A.~V.,  Panek R.~J.,   Schiffer Iii F.~H.,  1982, \mn@doi [The Astrophysical Journal] {10.1086/183714}, 252, L35

\bibitem[\protect\citeauthoryear{Honeycutt, Kafka  \& Robertson}{Honeycutt et~al.}{2013}]{honeycutt_wind_2013}
Honeycutt R.~K.,  Kafka S.,   Robertson J.~W.,  2013, \mn@doi [The Astronomical Journal] {10.1088/0004-6256/145/2/45}, 145, 45

\bibitem[\protect\citeauthoryear{Horne \& Marsh}{Horne \& Marsh}{1986}]{horne_emission_1986}
Horne K.,  Marsh T.~R.,  1986, \mn@doi [Monthly Notices of the Royal Astronomical Society] {10.1093/mnras/218.4.761}, 218, 761

\bibitem[\protect\citeauthoryear{Hunter}{Hunter}{2007}]{hunter_matplotlib_2007}
Hunter J.~D.,  2007, \mn@doi [Computing in Science \& Engineering] {10.1109/MCSE.2007.55}, 9, 90

\bibitem[\protect\citeauthoryear{Knigge \& Drew}{Knigge \& Drew}{1997}]{knigge_eclipse_1997}
Knigge C.,  Drew J.~E.,  1997, \mn@doi [The Astrophysical Journal] {10.1086/304519}, 486, 445

\bibitem[\protect\citeauthoryear{Knigge, Woods  \& Drew}{Knigge et~al.}{1995}]{knigge_application_1995}
Knigge C.,  Woods J.~A.,   Drew J.~E.,  1995, \mn@doi [Monthly Notices of the Royal Astronomical Society] {10.1093/mnras/273.2.225}, 273, 225

\bibitem[\protect\citeauthoryear{Knigge, Long, Blair  \& Wade}{Knigge et~al.}{1997}]{knigge_disks_1997}
Knigge C.,  Long K.~S.,  Blair W.~P.,   Wade R.~A.,  1997, \mn@doi [The Astrophysical Journal] {10.1086/303607}, 476, 291

\bibitem[\protect\citeauthoryear{La~Dous}{La~Dous}{1990}]{la_dous_catalogue_1990}
La~Dous C.,  1990, \mn@doi [Space Science Reviews] {10.1007/BF00167124}, 52, 203

\bibitem[\protect\citeauthoryear{Leighly \& Moore}{Leighly \& Moore}{2004}]{leighly_hst_2004}
Leighly K.~M.,  Moore J.~R.,  2004, \mn@doi [The Astrophysical Journal] {10.1086/422088}, 611, 107

\bibitem[\protect\citeauthoryear{Linnell, Szkody, Gänsicke, Long, Sion, Hoard  \& Hubeny}{Linnell et~al.}{2005}]{linnell_mv_2005}
Linnell A.~P.,  Szkody P.,  Gänsicke B.,  Long K.~S.,  Sion E.~M.,  Hoard D.~W.,   Hubeny I.,  2005, \mn@doi [The Astrophysical Journal] {10.1086/429143}, 624, 923

\bibitem[\protect\citeauthoryear{Mata~Sánchez et~al.,}{Mata~Sánchez et~al.}{2018}]{mata_sanchez_1989_2018}
Mata~Sánchez D.,  et~al., 2018, \mn@doi [Monthly Notices of the Royal Astronomical Society] {10.1093/mnras/sty2402}, 481, 2646

\bibitem[\protect\citeauthoryear{Mata~Sánchez et~al.,}{Mata~Sánchez et~al.}{2022}]{mata_sanchez_hard-state_2022}
Mata~Sánchez D.,  et~al., 2022, Hard-state optical wind during the discovery outburst of the black-hole {X}-ray dipper {MAXI} {J1803}-298, \mn@doi{10.48550/arXiv.2201.09896}, \url {http://arxiv.org/abs/2201.09896}

\bibitem[\protect\citeauthoryear{Mata~Sánchez, Muñoz-Darias, Casares, Huertas-Company  \& Panizo-Espinar}{Mata~Sánchez et~al.}{2023}]{mata_sanchez_ask_2023}
Mata~Sánchez D.,  Muñoz-Darias T.,  Casares J.,  Huertas-Company M.,   Panizo-Espinar G.,  2023, \mn@doi [Monthly Notices of the Royal Astronomical Society] {10.1093/mnras/stad1895}, 524, 338

\bibitem[\protect\citeauthoryear{Matthews}{Matthews}{2016}]{matthews_disc_2016}
Matthews J.,  2016, phd, University of Southampton, \url {https://eprints.soton.ac.uk/400903/}

\bibitem[\protect\citeauthoryear{Matthews, Knigge, Long, Sim  \& Higginbottom}{Matthews et~al.}{2015}]{matthews_impact_2015}
Matthews J.~H.,  Knigge C.,  Long K.~S.,  Sim S.~A.,   Higginbottom N.,  2015, \mn@doi [Monthly Notices of the Royal Astronomical Society] {10.1093/mnras/stv867}, 450, 3331

\bibitem[\protect\citeauthoryear{Matthews, Knigge, Long, Sim, Higginbottom  \& Mangham}{Matthews et~al.}{2016}]{matthews_testing_2016}
Matthews J.~H.,  Knigge C.,  Long K.~S.,  Sim S.~A.,  Higginbottom N.,   Mangham S.~W.,  2016, \mn@doi [Monthly Notices of the Royal Astronomical Society] {10.1093/mnras/stw323}, 458, 293

\bibitem[\protect\citeauthoryear{Matthews, Knigge, Higginbottom, Long, Sim, Mangham, Parkinson  \& Hewitt}{Matthews et~al.}{2020}]{matthews_stratified_2020}
Matthews J.~H.,  Knigge C.,  Higginbottom N.,  Long K.~S.,  Sim S.~A.,  Mangham S.~W.,  Parkinson E.~J.,   Hewitt H.~A.,  2020, \mn@doi [Monthly Notices of the Royal Astronomical Society] {10.1093/mnras/staa136}, 492, 5540

\bibitem[\protect\citeauthoryear{Matthews et~al.,}{Matthews et~al.}{2023}]{matthews_disc_2023}
Matthews J.~H.,  et~al., 2023, \mn@doi [Monthly Notices of the Royal Astronomical Society] {10.1093/mnras/stad2895}, 526, 3967

\bibitem[\protect\citeauthoryear{Matthews et~al.,}{Matthews et~al.}{2024}]{matthews_sirocco_2024}
Matthews J.~H.,  et~al., 2024, {SIROCCO}: {A} {Publicly} {Available} {Monte} {Carlo} {Ionization} and {Radiative} {Transfer} {Code} for {Astrophysical} {Outflows}, \url {http://arxiv.org/abs/2410.19908}

\bibitem[\protect\citeauthoryear{McGraw, Shields, Hamann, Capellupo  \& Herbst}{McGraw et~al.}{2018}]{mcgraw_quasar_2018}
McGraw S.~M.,  Shields J.~C.,  Hamann F.~W.,  Capellupo D.~M.,   Herbst H.,  2018, \mn@doi [Monthly Notices of the Royal Astronomical Society] {10.1093/mnras/stx3219}, 475, 585

\bibitem[\protect\citeauthoryear{Miller, Raymond, Reynolds, Fabian, Kallman  \& Homan}{Miller et~al.}{2008}]{miller_accretion_2008}
Miller J.~M.,  Raymond J.,  Reynolds C.~S.,  Fabian A.~C.,  Kallman T.~R.,   Homan J.,  2008, \mn@doi [The Astrophysical Journal] {10.1086/588521}, 680, 1359

\bibitem[\protect\citeauthoryear{Muñoz-Darias et~al.,}{Muñoz-Darias et~al.}{2019}]{munoz-darias_hard-state_2019}
Muñoz-Darias T.,  et~al., 2019, \mn@doi [The Astrophysical Journal Letters] {10.3847/2041-8213/ab2768}, 879, L4

\bibitem[\protect\citeauthoryear{Muñoz-Darias et~al.,}{Muñoz-Darias et~al.}{2020}]{munoz-darias_changing-look_2020}
Muñoz-Darias T.,  et~al., 2020, \mn@doi [The Astrophysical Journal] {10.3847/2041-8213/ab8381}, 893, L19

\bibitem[\protect\citeauthoryear{Panizo-Espinar, Muñoz-Darias, Armas~Padilla, Jiménez-Ibarra, Casares  \& Mata~Sánchez}{Panizo-Espinar et~al.}{2021}]{panizo-espinar_optical_2021}
Panizo-Espinar G.,  Muñoz-Darias T.,  Armas~Padilla M.,  Jiménez-Ibarra F.,  Casares J.,   Mata~Sánchez D.,  2021, \mn@doi [Astronomy \& Astrophysics] {10.1051/0004-6361/202140323}, 650, A135

\bibitem[\protect\citeauthoryear{Parkinson, Knigge, Dai, Thomsen, Matthews  \& Long}{Parkinson et~al.}{2024}]{parkinson_multi-dimensional_2024}
Parkinson E.~J.,  Knigge C.,  Dai L.,  Thomsen L.~L.,  Matthews J.~H.,   Long K.~S.,  2024, A multi-dimensional view of a unified model for {TDEs}, \mn@doi{10.48550/arXiv.2408.16371}, \url {https://ui.adsabs.harvard.edu/abs/2024arXiv240816371P}

\bibitem[\protect\citeauthoryear{Patterson, Thorstensen, Fried, Skillman, Cook  \& Jensen}{Patterson et~al.}{2001}]{patterson_superhumps_2001}
Patterson J.,  Thorstensen J.~R.,  Fried R.,  Skillman D.~R.,  Cook L.~M.,   Jensen L.,  2001, \mn@doi [Publications of the Astronomical Society of the Pacific] {10.1086/317973}, 113, 72

\bibitem[\protect\citeauthoryear{Pedregosa et~al.,}{Pedregosa et~al.}{2011}]{pedregosa_scikit-learn_2011}
Pedregosa F.,  et~al., 2011, Journal of Machine Learning Research, 12, 2825

\bibitem[\protect\citeauthoryear{Prinja \& Rosen}{Prinja \& Rosen}{1995}]{prinja_high-resolution_1995}
Prinja R.~K.,  Rosen R.,  1995, \mn@doi [Monthly Notices of the Royal Astronomical Society] {10.1093/mnras/273.2.461}, 273, 461

\bibitem[\protect\citeauthoryear{Prinja, Long, Froning, Knigge, Witherick, Clark  \& Ringwald}{Prinja et~al.}{2003}]{prinja_fuse_2003}
Prinja R.~K.,  Long K.~S.,  Froning C.~S.,  Knigge C.,  Witherick D.~K.,  Clark J.~S.,   Ringwald F.~A.,  2003, \mn@doi [Monthly Notices of the Royal Astronomical Society] {10.1046/j.1365-8711.2003.06307.x}, 340, 551

\bibitem[\protect\citeauthoryear{Richards, Vanden~Berk, Reichard, Hall, Schneider, SubbaRao, Thakar  \& York}{Richards et~al.}{2002}]{richards_broad_2002}
Richards G.~T.,  Vanden~Berk D.~E.,  Reichard T.~A.,  Hall P.~B.,  Schneider D.~P.,  SubbaRao M.,  Thakar A.~R.,   York D.~G.,  2002, \mn@doi [The Astronomical Journal] {10.1086/341167}, 124, 1

\bibitem[\protect\citeauthoryear{Ringwald \& Naylor}{Ringwald \& Naylor}{1998}]{ringwald_high-speed_1998}
Ringwald F.~A.,  Naylor T.,  1998, \mn@doi [The Astronomical Journal] {10.1086/300192}, 115, 286

\bibitem[\protect\citeauthoryear{Ritter \& Kolb}{Ritter \& Kolb}{2003}]{ritter_catalogue_2003}
Ritter H.,  Kolb U.,  2003, \mn@doi [Astronomy \& Astrophysics] {10.1051/0004-6361:20030330}, 404, 301

\bibitem[\protect\citeauthoryear{Scepi, Lesur, Dubus  \& Flock}{Scepi et~al.}{2018}]{scepi_turbulent_2018}
Scepi N.,  Lesur G.,  Dubus G.,   Flock M.,  2018, \mn@doi [Astronomy and Astrophysics] {10.1051/0004-6361/201833921}, 620, A49

\bibitem[\protect\citeauthoryear{Segura et~al.,}{Segura et~al.}{2022}]{segura_persistent_2022}
Segura N.~C.,  et~al., 2022, \mn@doi [Nature] {10.1038/s41586-021-04324-2}, 603, 52

\bibitem[\protect\citeauthoryear{Skillman, Patterson  \& Thorstensen}{Skillman et~al.}{1995}]{skillman_superhumps_1995}
Skillman D.~R.,  Patterson J.,   Thorstensen J.~R.,  1995, \mn@doi [Publications of the Astronomical Society of the Pacific] {10.1086/133590}, 107, 545

\bibitem[\protect\citeauthoryear{Smak}{Smak}{1981}]{smak_emission_1981}
Smak J.,  1981, Acta Astronomica, 31, 395

\bibitem[\protect\citeauthoryear{Tampo, Knigge, Long, Matthews  \& Segura}{Tampo et~al.}{2024}]{tampo_disc_2024}
Tampo Y.,  Knigge C.,  Long K.~S.,  Matthews J.~H.,   Segura N.~C.,  2024, A disc wind origin for the optical spectra of dwarf novae in outburst, \mn@doi{10.48550/arXiv.2406.14396}, \url {http://arxiv.org/abs/2406.14396}

\bibitem[\protect\citeauthoryear{Trueba, Miller, Kaastra, Zoghbi, Kallman, Proga  \& Raymond}{Trueba et~al.}{2019}]{trueba_comprehensive_2019}
Trueba N.,  Miller J.~M.,  Kaastra J.,  Zoghbi A.,  Kallman A. C. F.~T.,  Proga D.,   Raymond J.,  2019, \mn@doi [The Astrophysical Journal] {10.3847/1538-4357/ab4f70}, 886, 104

\bibitem[\protect\citeauthoryear{Virtanen et~al.,}{Virtanen et~al.}{2020}]{virtanen_scipy_2020}
Virtanen P.,  et~al., 2020, \mn@doi [Nature Methods] {10.1038/s41592-019-0686-2}, 17, 261

\bibitem[\protect\citeauthoryear{Vitello \& Shlosman}{Vitello \& Shlosman}{1993}]{vitello_ultraviolet_1993}
Vitello P.,  Shlosman I.,  1993, \mn@doi [The Astrophysical Journal] {10.1086/172799}, 410, 815

\bibitem[\protect\citeauthoryear{Weymann, Morris, Foltz  \& Hewett}{Weymann et~al.}{1991}]{weymann_comparisons_1991}
Weymann R.~J.,  Morris S.~L.,  Foltz C.~B.,   Hewett P.~C.,  1991, \mn@doi [The Astrophysical Journal] {10.1086/170020}, 373, 23

\bibitem[\protect\citeauthoryear{Zhao, Wang  \& Liu}{Zhao et~al.}{2025}]{zhao_searching_2025}
Zhao X.,  Wang S.,   Liu J.,  2025, Searching for accreting compact binary systems from spectroscopy and photometry: {Application} to {LAMOST} spectra, \mn@doi{10.48550/arXiv.2503.12410}, \url {http://arxiv.org/abs/2503.12410}

\makeatother
\end{thebibliography}



\appendix

\section{A Physically Motivated Approximation For the Luminosity of Wind-Formed Recombination Lines}
\label{sec:CoG Appendix}

\begin{table}
\centering
\begin{tabular}{|c|c|c|c|}
\hline
 (i)& \multicolumn{3}{| c |}{Using True Emission Measure} \\
\hline
Inclination (°) & $\mathrm{K_1}$ & $\mathrm{K_2}$ & $\mathrm{Scatter [\sigma] (dex)}$\\
\hline
20 & $1.04\times10^{33}$ & $-55.41$ & $0.31$   \\
\hline
45 & $8.82\times10^{32}$ & $-55.52$ & $0.35$  \\
\hline
60 & $8.09\times10^{32}$ & $-55.59$ & $0.39$  \\
\hline
72.5 & $7.45\times10^{32}$ & $-55.62$ & $0.44$  \\
\hline
85  & $5.87\times10^{32}$ & $-55.33$ & $0.45$\\
\hline
\hline
(ii)& \multicolumn{3}{| c |}{Using Predicted Emission Measure}\\
\hline
Inclination (°) & $\mathrm{K_1}$ & $\mathrm{K_2}$ & $\mathrm{Scatter [\sigma] (dex)}$\\
\hline
20  &$1.24\times10^{33}$ & $55.72$ & $0.33$  \\
\hline
45 & $1.52\times10^{33}$ & $-56.18$ & $0.42$  \\
\hline
60 & $1.52\times10^{33}$ & $-56.25$ & $0.45$ \\
\hline
72.5 & $2.02\times10^{33}$ & $-55.62$ & $0.55$ \\
\hline
85  & $1.54\times10^{33}$ & $-56.20$ & $0.53$\\
\hline
\end{tabular}
\caption{The constants $K_1$ and $K_2$ in equation \ref{fitting CoG} as determined by simple least squares fits. Results are shown for two cases: (i) when using the {\em actual} emission measures from the \textsc{sirocco} models for fitting equation \ref{fitting CoG}; (ii) when the emission measure is {\em estimated} from equation \ref{eq:scaling_EM} using the wind input parameters. The RMS scatters (in dex) around the fits are also provided. All fits are to the full data set, i.e. Gold, Silver and Bronze (so long as $\mathrm{EW > 0\,\mathring{A}}$).}
\label{tab:CoG_table}
\end{table}

The luminosity of an emission line formed in an optically thin medium is directly proportional to the volumetric emission measure, 
\begin{equation}
\label{equ:EM equation}
L_{line} \propto EM = \int_{wind} N^2_e dV.
\end{equation}
Here, $N_e$ is the number density of free electrons and the integration is over the entire emitting volume \citep[e.g.][]{frank_accretion_2002,dennis_emission_2024}. We can calculate the emission measures associated with each of our models trivially, as \textsc{sirocco} records the electron number density and cell volume across our outflowing structure. 

Once the line-of-sight optical depth in the line-forming region reaches $\tau \simeq 1$, the line emission begins to saturate and the EW will no longer increase linearly with EM. For a recombination line formed in a uniform cylindrical slab of radius $R$, height $H$ and electron density $N_e$ (i.e. the base of the accretion disc wind), one can find that $\tau \propto EM/R^2$. Since the accretion disc radius is fixed in all of our models, $\tau \propto EM$ to zeroth order. 

Suppose we approximate the line profile as Gaussian. In that case, we can follow the standard curve-of-growth (CoG) analysis to estimate how the line luminosity may scale with $\tau$ (and hence EM)  as the line becomes optically thick. The result is 
\begin{equation}
    L_{line} = K_1 \int_{0}^{\infty} [1-exp(K_2\times EM e^{-x^2})]\,dx \
    \label{fitting CoG}
\end{equation}
where $K_1$ and $K_2 = \tau / EM$ are constants. The integral is over the line profile with $x$ representing velocity in units of the Doppler width of the line. In the optically thin limit, $K_2 \times EM << 1$, this equation reduces to $L_{line} \propto EM$. In the optically thick limit, line photons escape primarily in the line wings (which remain optically thin). The CoG then turns over and flattens, scaling as $L_{line} \propto \ln(\tau) \propto \ln(EM)$.

\begin{figure*}
    \centering
    \includegraphics[width=0.85\linewidth]{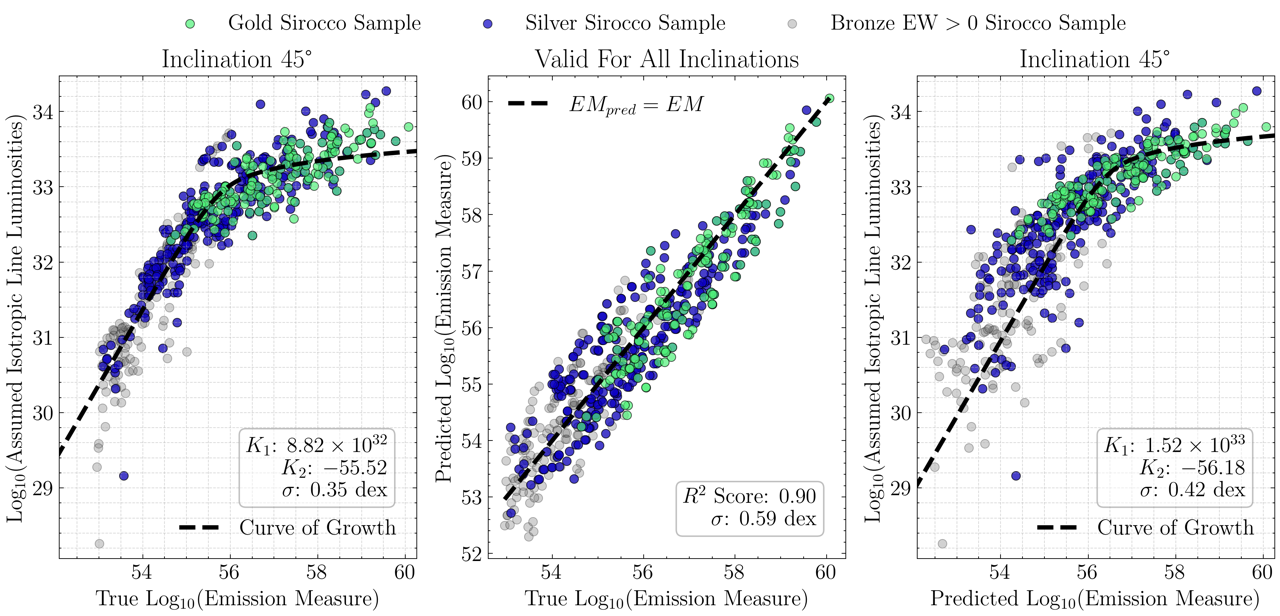}
    \caption{The left plot displays the underlying relation that connects the emission measure of a disc wind to the $\mathrm{H\,\alpha}$ line luminosity. The plot is for a 45° inclination. However, all other inclinations follow an identical shape but with different curve-of-growth fitting parameters. The middle plot shows the predictive performance of the emission measure scaling relation given an outflow's particular combination of \textsc{sirocco} parameters. The emission measure is calculated from the disc wind model and, therefore, valid for all inclinations and samples. The black dashed $y=x$ line highlights an ideal 1:1 scenario where the predicted emission measure exactly matches the calculated emission measure. The right plot, similar to the left, shows the $\mathrm{H\,\alpha}$ line luminosity but now utilising the predicted emission measure from the scaling relation. Green data points highlight the Gold sample, navy points highlight the Silver sample and light grey points highlight all the Bronze sample with positive EW spectra. We present numeric values for all inclination-dependent parameter coefficients in Table \ref{tab:CoG_table}.} 
    \label{fig:appendix_CoG}
\end{figure*}

The left-hand panel of Fig. \ref{fig:appendix_CoG} shows how the $\mathrm{H\,\alpha}$ line luminosities in our models scale with EM for one particular inclination. The dashed line is a least-squares fit to these points based on equation \ref{fitting CoG}, with $K_1$ and $K_2$ treated as free parameters. This simple, physically motivated model represents the data reasonably well and suggests that the line profiles in our Gold sample (and hence wind-formed lines of observed CVs) are formed under at least marginally optically thick conditions. Table \ref{tab:CoG_table} (i) lists these fit parameters $K_1$ and $K_2$, as well as the RMS scatter around the fits for all inclinations. 

In order to relate the luminosity of a line to the physical outflow properties, we can try to approximate the emission measure as (mainly) a power law in the disc wind parameters. Such an approximation is illustrated in the middle panel of Fig. \ref{fig:appendix_CoG} and is given analytically by
\begin{equation}
\begin{split}
\label{eq:scaling_EM}
\mathrm{EM \simeq 10^{54.5} cm^{-3}} ~ & \left[\frac{\dot{M}_{acc}}{\mathrm{10^{-8}~M_{\odot}~yr^{-1}}}\right]^{0.00}
~\left[\frac{\dot{M}_{wind}}{\mathrm{10^{-9}~M_{\odot}~yr^{-1}}}\right]^{1.96}\\
& \left[\frac{d}{\mathrm{5.5~R_{WD}}}\right]^{0.21}
~\left[\frac{R_s}{\mathrm{10~R_{WD}}}\right]^{1.14}\\
& \left[10^{0.7(\alpha-0.25)}\right]
~ \left[10^{0.6(\beta-1.5)}\right].
\end{split}
\end{equation}

We can then refit the line luminosities with equation \ref{fitting CoG}, using the approximate emission measure estimates provided by equation \ref{eq:scaling_EM}. This fit obviously yields slightly different values for $K_1$ and $K_2$ and is shown in the right-hand panel of Fig. \ref{fig:appendix_CoG}. However, now, advantageously, this figure is based on \textsc{sirocco} wind modelling parameters. Table \ref{tab:CoG_table} (ii) lists these new fit parameters and the corresponding RMS scatter around these fits.

\section{Reference diagnostic diagrams}

\begin{figure*}
    \centering
    \includegraphics[width=0.85\textwidth]{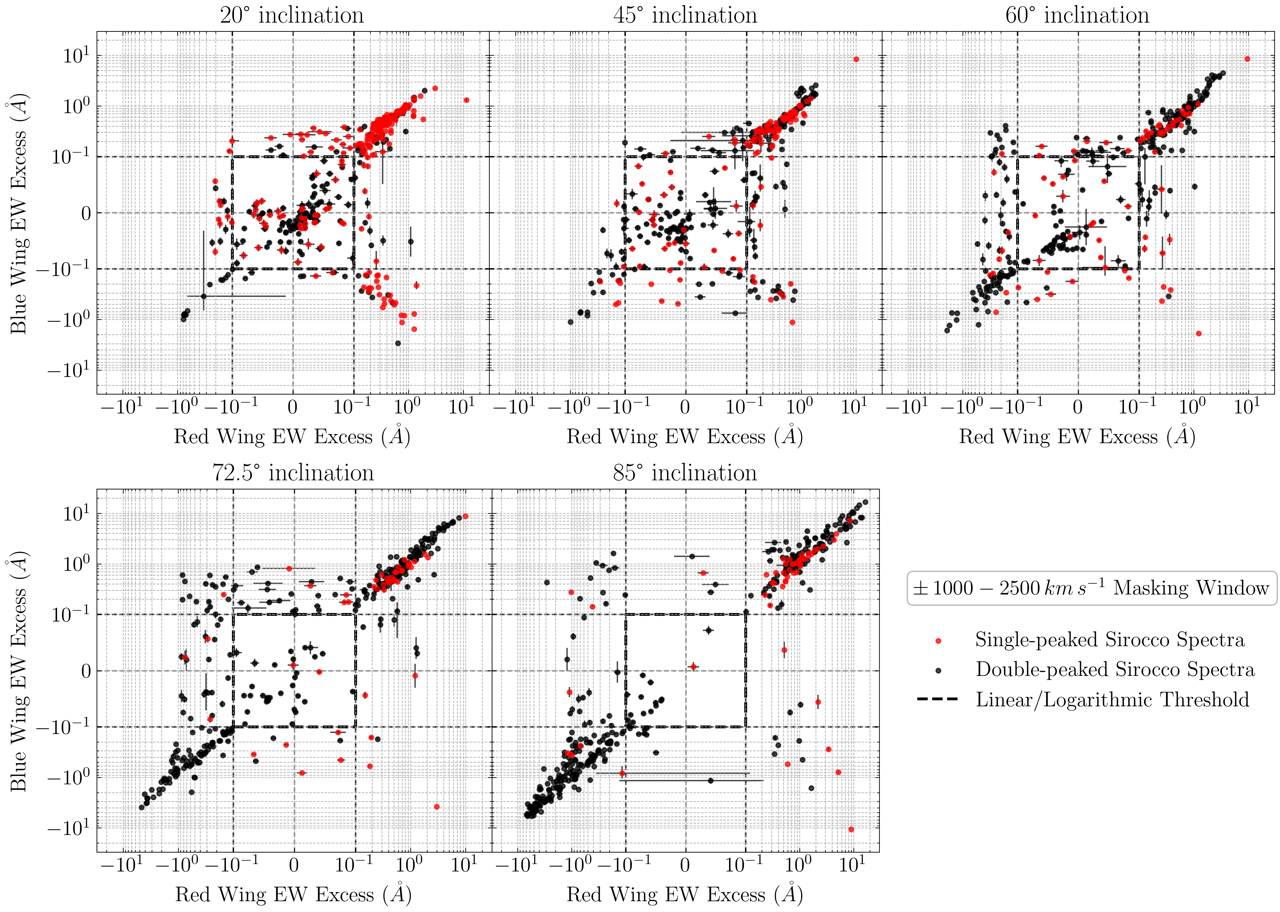}
    \caption{Excess diagnostic diagrams for Silver \textsc{sirocco} CV systems viewed at a 20°, 45°, 60°, 72.5° and 85° inclination. Red data points indicate \textsc{sirocco} spectra that exhibit only one prominent peak within the $\mathrm{H\,\alpha}$ emission line. Black data points indicate \textsc{sirocco} spectra with two prominent peaks. The masking profile is set to a radial velocity at $\pm\,\mathrm{1000-2500\,km\,s^{-1}}$ which corresponds to wavelengths $\pm\,\mathrm{22-55\,\mathring{A}}$ from the rest wavelength of $\mathrm{H\,\alpha}$ ($\mathrm{6562.819\,\mathring{A}}$). The linear-logarithmic threshold for the symmetric logarithmic scale plot, shown as dashed black lines, is set at $\mathrm{0.1\,\mathring{A}}$.}
    \label{fig:appendix_diag_plots_22_55}
\end{figure*}

\begin{figure*}
    \centering
    \includegraphics[width=0.85\textwidth]{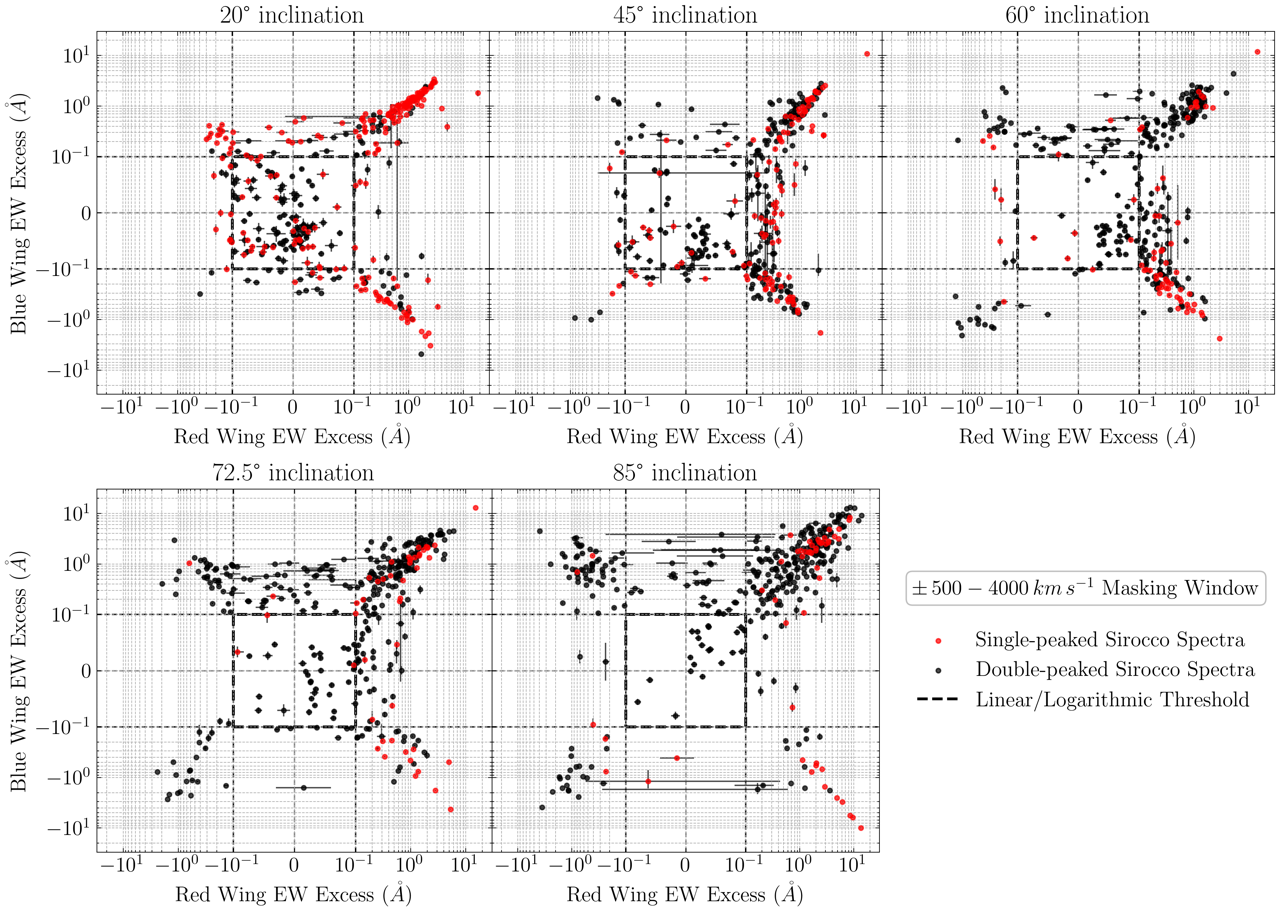}
    \caption{Same as Fig. \ref{fig:appendix_diag_plots_22_55} with fixed masking window set to a radial velocity at $\pm\,\mathrm{500-4000\,km\,s^{-1}}$ ($\pm\,\mathrm{11-88\,\mathring{A}}$) from $\mathrm{H\,\alpha}$'s rest wavelength.}
    \label{fig:appendix_diag_plots_11_88}
\end{figure*}

\begin{figure*}
    \centering
    \includegraphics[width=0.85\linewidth]{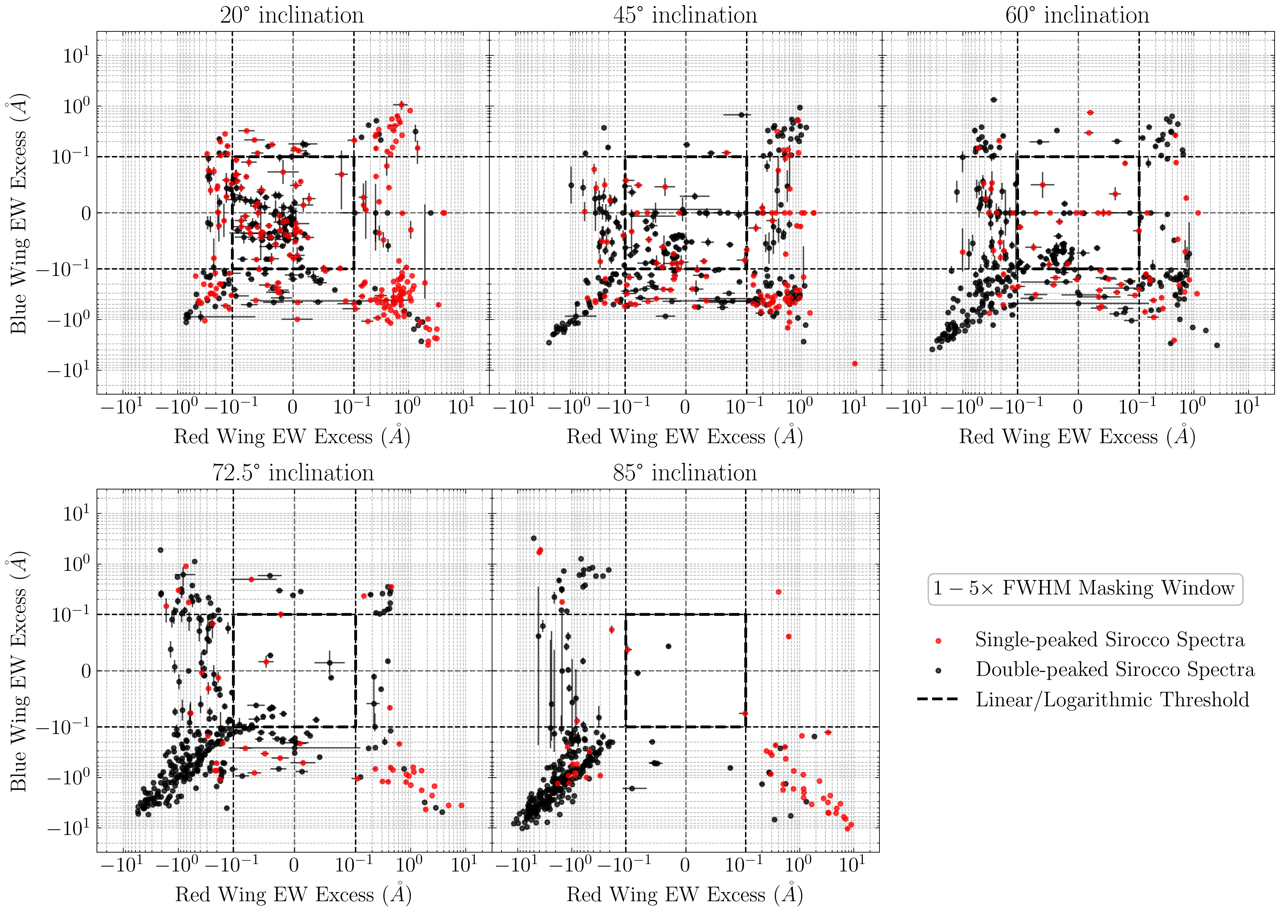}
    \caption{Same as Fig. \ref{fig:appendix_diag_plots_22_55} with dynamic masking window set at $\mathrm{1.0\times}$ and $\mathrm{5.0\times}$ of a line profile's respective FWHM from $\mathrm{H\,\alpha}$'s rest wavelength.}
    \label{fig:appendix_diag_plots_fwhm}
\end{figure*}

Here, we provide another three excess EW diagnostic diagrams, similar to those seen previously, but now including the Silver sample for all inclinations. Fig. \ref{fig:appendix_diag_plots_22_55} utilises the $\mathrm{\pm\, 1000-2500\,km\,s^{-1}}$ masking profile adopted throughout this study. Fig. \ref{fig:appendix_diag_plots_11_88} showcases the impact severity when adopting a wider mask at $\mathrm{\pm\, 500-4000\,km\,s^{-1}}$. Fig. \ref{fig:appendix_diag_plots_fwhm} displays a new spread of data across the diagram by using our new FWHM-based masking approach. Remember that data points falling into particular regions of the diagnostic diagram are intended to predict a particular line profile shape. These predictions are more reliable when the spectrum is single-peaked. 

\label{lastpage}
\end{document}